\documentclass{article}

\usepackage{arxiv}
\usepackage{tikz}
\usepackage[utf8]{inputenc} 
\usepackage[T1]{fontenc}    
\usepackage{hyperref}       
\usepackage{url}            
\usepackage{booktabs}       
\usepackage{paralist}
\usepackage{amssymb}
\usepackage{amsfonts}       
\usepackage{nicefrac}       
\usepackage{microtype}      
\usepackage{lipsum}		
\usepackage{graphicx}
\usepackage{natbib}
\usepackage{doi,xcolor}
\graphicspath{{./images}}
\usepackage{caption,subcaption}
\usepackage{amsmath,bm}
\usepackage{appendix}

\newcommand{\p}{\mathbf{p}}  

\newcommand{\dz}{d^z} 
\newcommand{\Dz}{D^z}
\newcommand{\dr}{d^r} 
\newcommand{\Dr}{D^r}
\newcommand{\gz}{g^z} 
\newcommand{\Gz}{G^z}
\newcommand{\gr}{g^r} 
\newcommand{\Gr}{G^r}
\newcommand{\Grz}{G^{r,z}}
\newcommand{\grz}{g^{r,z}}
\newcommand{\nz}{\epsilon^z} 
\newcommand{\Nz}{\varepsilon^z}
\newcommand{\nr}{\epsilon^r} 
\newcommand{\Nr}{\varepsilon^r}
\newcommand{\Nt}{\varepsilon}
\newcommand{\nt}{\epsilon}

\newcommand{\rec}{\mathbf{r}} 
\newcommand{\coh}{\mathbf{q}}           
\newcommand{\nuis}{\mathbf{p}}        
\newcommand{\vir}{\mathbf{y}} 

\newcommand{\qv}{\bm{\epsilon}}  

\newcommand{\qln}{q}  

\newcommand{\post}{Q} 
\newcommand{\likl}{P} 
\newcommand{\Ncode}{\hat{\mathbf{m}}} 
\newcommand{\data}{\mathbf{x}} 
\newcommand{\lat}{\mathbf{z}} 
\newcommand{\n}{n} 
\newcommand{\Nn}{N} 
\newcommand{\ncc}{\gamma}
\newcommand{\drec}{\mathbf{r}^{\prime}_j} 
\newcommand{\rrec}{\mathbf{r}^{\text{true}}_j} %
\newcommand{\mcoh}{\bar{\coh}} 
\newcommand{\mnuis}{\bar{\nuis}} 

\title{Enhanced Receiver Function imaging of crustal structures using Symmetric Autoencoders}

\author{ \href{https://orcid.org/0009-0007-2912-8242}{\includegraphics[scale=0.06]{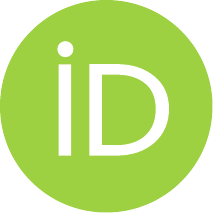}\hspace{1mm}T. Rengneichuong Koireng} \\
	Centre for Earth Sciences\\
	Indian Institute of Science\\
	Bengaluru, India 560012 \\
	\texttt{tienterk@iisc.ac.in} \\
	\And
	\href{https://orcid.org/0000-0003-4081-8969}{\includegraphics[scale=0.06]{orcid.pdf}\hspace{1mm}Pawan Bharadwaj} \\
	Centre for Earth Sciences\\
	Indian Institute of Science\\
	Bengaluru, India 560012 \\
	\texttt{pawan@iisc.ac.in} \\
}

\renewcommand{\shorttitle}{Enhanced RF Imaging of Crustal Structures Using SymAE}

\hypersetup{
pdftitle={Enhanced Receiver Function Imaging Of Crustal Structures Using Symmetric Autoencoders},
pdfauthor={T~Rengneichuong~Koireng, Pawan~Bharadwaj},
pdfkeywords={Receiver functions, Autoencoders, Coherent, Deconvolution, Crustal Imaging, Subduction},
}

\begin{document}

\maketitle

\begin{abstract}

 The receiver-function (RF) technique aims to recover receiver-side crustal and mantle structures by deconvolving either the radial or transverse component with the vertical component seismogram. Analysis of the variations of RFs along the backazimuth and slowness is the key in determining the geometry and anisotropic properties of the crustal structures. However, the deconvolution introduces pseudorandom nuisance effects, due to unknown earthquake source signatures and seismic noise, which obstruct the precise extraction of backazimuth and slowness dependent crustal effects. Our goal is to obtain RFs with minimal nuisance effects, while preserving the crustal effects. In this study, we introduced a new method for reducing nuisance effects in RFs. This method generates virtual RFs through a deep generative model, namely symmetric variational autoencoders (SymVAE). Our autoencoder efficiently learns to disentangle coherent crustal effects and nuisance effects within its latent space, given a set of RFs derived from a cluster of nearby earthquakes. This disentanglement enables generation of virtual RFs which exhibits minimal nuisance effects while preserving the coherent crustal effects.
  We tested SymVAE using synthetic RFs with ambient seismic noise. We also tested using dense seismic networks in two distinct geological settings: the Cascadia subduction zone and southern California. We compared our method with linear and phase-weighted averaging. In both synthetic and real RFs, the generated virtual RFs demonstrate enhanced information related to crustal structures. We have also quantitatively assessed the performance. One major advantage of our method over traditional methods is its ability to utilize all available earthquake data, regardless of signal quality, resulting in improved backazimuth and slowness coverage.


\end{abstract}


\section{Introduction}

Receiver functions (RFs), computed using several teleseismic earthquakes, provide a powerful method for investigating crustal and upper mantle structures below a particular receiver. 
This study focuses on analyzing the changes in the RF relative to the backazimuth and the distance to the teleseismic earthquake, which are essential for accurate subsurface characterization. Careful analysis of these variations and comparison with forward modeling results~\citep{levin2008seismic, Savage2007, liu2015crustal} can reveal important information about the orientation and magnitude of Earth's anisotropy, as well as the 3-D geometry of subsurface features, providing researchers with valuable insights into the tectonic evolution and dynamics of a given region.
To illustrate this emphasis, we consider the systematic variation of the PS phase (forward-scattered S waves from the Moho caused by an incident P wave) with respect to backazimuth and distance.
Numerous studies have extracted the fast-polarization direction and magnitude of crustal anisotropy from the arrival of PS within RFs~\citep{McNamara1993, Bianchi2010, Zheng2018}. 
Similarly, 
backazimuth variations of the PS arrival in both radial and transverse RFs have been employed to calculate the dip azimuth, dip angle, and depth of subducting slabs~\citep{ShiomiK}. 
~\cite{zhu2000moho} employed RFs generated by earthquakes that occur at various epicentral distances to determine the thickness of the crust and the P to S velocity ratio through semblance analysis.


To analyze RF variations with respect to backazimuth and epicentral distance, one must be able to compare RFs from multiple earthquakes. However, this task can be challenging due to the presence of pseudo-random \emph{nuisance} effects.
The nuisance effects are unique to each earthquake and it should be emphasized that the common assumption that these effects can be modeled as random Gaussian noise frequently proves inadequate~\citep{kolb2014receiver,Bodin2014Inversion,bona1998variance}.
To be precise, 
nuisance effects are unwanted signals in RFs that arise through two distinct pathways during the deconvolution of the radial with the vertical component of the seismogram. First, as deconvolution is ill-posed, it is susceptible to contamination by background seismic noise (owing to natural and human
activities). 
This means that insufficient regularization in deconvolution amplifies the seismic noise linked to the spectral zeros of the source signature, which causes nuisance effects~\citep{Akuhara2019}. Conversely, when regularizing the deconvolution, nuisance effects appear as biases that are specific to the regularization method.
Secondly, deconvolution, which is associated with cross-correlation, results in crosstalk between the scattered waves from crustal structures and seismic noise in the vertical and radial components.
Crosstalk increases interference and makes the interpretation of converted seismic phases related to crustal structures (referred to as crustal effects) more challenging.
%
%

Various researchers~\citep{Park2000,Zhang2022,zhang2024crustal} have recognized that pseudo-random nuisance effects make it difficult to interpret receiver functions, highlighting the need to address these effects to achieve the accurate extraction of crustal effects.
A strategy to improve the quality of RFs involves careful selection and identification of usable earthquakes through automated processes~\citep{Crotwell2005,Yang2016} and machine learning frameworks~\cite{Gong2022,Krueger2021,sabermahani2024deepRFQC}. However, removing earthquakes will decrease azimuthal coverage. 
Therefore, most of the methods focus on reducing nuisance effects\footnote{The nuisance effects cannot be effectively removed through standard bandpass filtering techniques, as they share a frequency band similar to that of crustal converted phases.} in RFs. 
In the literature, these methods can be classified into two categories.

The first category corresponds to data-driven methods, where multiple RFs are averaged to improve the signal-to-noise ratio (SNR). Examples include
\begin{enumerate}
    \item averaging multiple earthquakes with a similar backazimuth and epicentral distance~\citep{Gurrola1995,Levin1997,Hu2015,Bloch2023};
    \item weighted averaging techniques, such as the frequency-domain method proposed by ~\cite{Park2000}, which can assess the level of nuisance effects in a RF.
    \item phase-weighted averaging~\citep{Schimmel1997} of multiple RFs with similar backazimuth and epicentral distance.
\end{enumerate}
The methods in this category
are efficient when
dealing with 1) permanent stations that record numerous earthquakes;
2) a uniform earthquake distribution across backazimuth and epicentral distance.
The averaging procedure in these methods assumes that nuisance effects follow a simple distribution, such as a Gaussian distribution, which is inaccurate.
In practice, the earthquake distribution is non-uniform, meaning that some azimuthal or epicentral-distance bins have more earthquakes than others, which can lead to less efficient averaging for certain bins~\citep{Levin1997,Ozakin2015,Bloch2023}.
The selection of weights in
weighted averaging methods can be subjective.



Model-based techniques comprise the second category of methods that aim to minimize nuisance effects. To improve RF quality, these methods rely on assumptions about the Earth's velocity structure.
They apply constraints to a gather of available RFs, ordered according to either backazimuth or epicentral distance. 
Examples include
\begin{enumerate}
    \item methods using sparsity-promoting Randon transforms~\citep{Olugboji2023}, curvelet transforms ~\citep{chen2019denoising}, and singular spectrum analysis~\citep{Dokht2016} assume a laterally homogeneous Earth model to filter linear events from the RF gather;
    \item supervised deep learning techniques~\citep{Wang2022} are trained on crustal velocity models drawn from the prior distribution to output crustal thickness and the P-to-S wave velocity ratio from the RFs; 
    \item common conversion point (CCP) averaging~\citep{Zhu2000} of the RF gather, 
    commonly used for imaging the Moho and mantle discontinuities, involves time-depth conversion based on a large-scale velocity model.
\end{enumerate}
Methods in this category are unsuccessful when applied to intricate geological environments such as subduction zones, which are different from the assumed (simpler) models. These methods also require dense spatial coverage of earthquake events, which is challenging for temporary seismic stations.
CCP averaging may result in artifacts~\citep{Zheng2014} due to mismatches between the assumed velocity model and the Earth's true model.
%
An important drawback of supervised deep learning is the difficulty in assessing its
performance on real seismic data, which originates from a distribution different from that assumed during training.

This paper aims to train neural networks on RFs from various earthquakes and then \emph{synthesize} new virtual RFs with minimal nuisance effects.
We demonstrate that these virtual RFs retain crustal effects, enabling the extraction of their dependence on backazimuth and epicentral distance.
In computer vision,
various approaches have been devised to generate new images by learning the distribution of the training images~\citep{Kingma2014,diffusion,Kobyzev_2021}. 
We extend the variational autoencoder (VAE) framework~\citep{Kingma2014,bishop2023deep}, a nonlinear latent variable model in which the latent variables capture key features of the training data.
VAE
can generate virtual data (such as images, audio, or seismograms) by first selecting samples in the low-dimensional latent space and then decoding these samples.

In geophysics, autoencoders have been used to identify and classify seismograms based on a low-dimensional representation of the seismograms~\citep{Valentine2012}.  VAE has been used to generate realistic ground motions corresponding to specific physical features such as earthquake magnitude, soil type, and source focal mechanism~\citep{NingXie2024,Fayaz2023}. 
The use of the generated virtual data is problem-specific, where in our case we are interested in generating RFs with minimal nuisance effects.
In order to do so, we need to disentangle
the nuisance effects from the crustal effects in the latent space --- traditional VAE architectures do not have this feature.
Recent studies have focused on learning the disentangled latent space within the VAE framework~\citep{Mathieu2016,burgess2018}. 
 In this study, we employ a variation of VAE called symmetric variational autoencoders~\citep[SymVAE]{bharadwaj2024vsymae} to learn a latent space where the crustal and nuisance effects are disentangled. 
 %

To validate our methodology, we applied it to RFs derived from synthetic seismograms with real earthquake source signatures and ambient seismic noise.
%
We considered a crustal model with an anisotropic layer. 
 The quality of virtual RFs was assessed by calculating the normalized correlation coefficient (NCC) with respect to the true RFs. Virtual RFs exhibited higher NCC compared to RFs obtained by linear and phase-weighted averaging. We have evaluated the effectiveness of our method in complex geological settings, such as
the Cascadia subduction zone and southern California. Virtual radial RFs from Cascadia clearly highlight the scattered S-waves produced by the velocity interfaces of the subducting oceanic slab, as reported by~\cite{Bloch2023}. Radial virtual RFs along a profile intersecting the San Andreas fault in southern California show sharp variations in converted S-wave delay time, indicating inhomogeneous crustal thickness. 
Finally, the virtual RFs showed spatial consistency with patterns similar to those of adjacent seismic stations, confirming their geological validity. 
%
In contrast to many previous studies that discard low SNR RFs, our method leverages all available earthquakes. Our approach requires 
fewer assumptions regarding Earth models and nuisance statistics.
It enables the generation of high-quality RFs to increase the coverage of the backazimuth and epicentral distance. Furthermore, like other deep learning techniques, our method is automated and scalable to various datasets and geographic areas.

The paper is structured as follows. We begin by detailing the grouping of RFs necessary for preparing a training dataset for our deep-generative model i.e. SymVAE. Next, we explain how SymVAE disentangles crustal effects from nuisance effects and generates nuisance-minimized RFs. We then detail synthetic experiments and compare SymVAE's performance with conventional averaging methods. Finally, our method is applied to real data to evaluate its performance in complex geological settings.

\section{Grouping Receiver Functions To Create Datapoints}
A datapoint is an element from the input dataset employed to train a generative model. 
In this paper, each datapoint refers to a group of RFs with a similar backazimuth angle and epicentral distance. For each datapoint, our method aims to extract the coherent crustal effects shared by the RFs as discussed below. 

 \subsection{Receiver functions}
The wavefield due to a teleseismic earthquake, incident on the crustal structure near a seismic station, can be approximated as a planewave characterized by a seismic wavelet $s$ and a propagation direction defined by the wavenumber vector $\p$.
The interaction of this planewave with the crustal structure results in measured waveforms at the station, which can be expressed as a convolution of the seismic wavelet with the Earth's impulse response:
\begin{equation}
  \dz(t,\p) = \int_{\tau}s(t-\tau)\gz(\tau,\p)\,\text{d}\tau+\nz(t), 
  \label{eqn:one}
\end{equation}
\begin{equation}
    \dr(t,\p) = \int_{\tau}s(t-\tau)\gr(\tau,\p)\,\text{d}\tau+\mathit{\nr(t)}.
\end{equation}
Here, time is indicated as $t$, the vertical and radial components of the measured seismogram are indicated by using $\dz$ and $\dr$, respectively, while $\nz$ and $\nr$ represent uncorrelated noise in these components.
The vertical and radial impulse responses of the crust beneath the seismic station, $\gz$ and $\gr$, are necessary for imaging~\citep{Tauzin2019}. 
The seismic wavelet $s(t)$ contains information on the time history of the earthquake and the propagation effects owing to the structure of the Earth from the source to the base of the crust. 
The generation of radial receiver function involves the deconvolution of the radial component with the vertical component to predominantly extract SV converted waves. 
As shown in Appendix~\ref{sec:appn1}, the radial receiver function can be written as
\begin{equation}
    r(t,\p)= \int_{\tau}s_{\text{a}}(t-\tau,\p)\grz(\tau,\p)\,\text{d}\tau + \nt(t,\p), 
    \label{eqn:rfn}
\end{equation}
where $s_{\text{a}}$ denotes a zero-phase signal convolved with $\grz$, which depicts the cross-correlation between $\gr$ and $\gz$.
%
%
The nuisance effects are due to 1) $s_{\text{a}}$ which blurs $\grz$, and 2) the additive term 
$\nt$ which arises from the cross-correlation between seismic noise ($\nr$ and $\nz$) and the crustal impulse responses ($\gz$ and $\gr$). 
The purpose of this paper is to extract crustal effects in $\grz$, while minimizing nuisance effects for both radial and transverse RF.
%


\subsection{Coherent crustal effects}
Let us examine how $r$ changes with the wavenumber vector $\p$ in Eq.~\ref{eqn:rfn}. 
These changes can be due to the functions $\grz$, $s_\text{a}$, or $\nt$.
We assume that the function $\grz$ changes smoothly with $\p$, in contrast to $s_{\text{a}}$ and $\nt$, which fluctuate rapidly since each $\p$ corresponds to a distinct earthquake with a unique source time function and seismic noise.
This allowed us to distinguish the smooth variations of $\grz$ attributed to the geometry and anisotropy of the crustal layers from the rapid fluctuations of nuisance terms $s_{\text{a}}$ and $\nt$.
Our assumption 
essentially means, when analyzing the RFs of teleseismic earthquakes situated closely together, the crustal effects of $\grz$ are \emph{coherent} across RFs as $s_\text{a}$ is a zero-phase signal. A numerical validation of this assumption is detailed in Appendix.~\ref{sec:coherency}  

\subsection{Grouping receiver functions of a single station}
\label{sec:group}
In practical scenarios, the function $r$ as described in Eq.~\ref{eqn:rfn} can be computed for discrete time points and wavenumber vectors.
A receiver function will be obtained for each wavenumber vector corresponding to an earthquake with a unique backazimuth and epicentral distance.
We sorted the RFs into groups by establishing bins on the basis of chosen intervals of backazimuth and epicentral distance. The smaller the intervals, the fewer RFs each group contains. 
Group assignment involves associating each unique wavenumber vector value with a specific bin, determined by the corresponding backazimuth and epicentral distance ranges.
In our notation, subscript signifies the particular bin associated with the RF. The superscript denotes the index corresponding to the specific earthquake whose seismogram is used to compute the RF. Additionally, vectors after time discretization are denoted by boldface symbols. According to this notation and Eq.~\ref{eqn:rfn}, the RF derived from the seismograms of the $i$-th earthquake from in the $j$-th bin, after discretization, is expressed in time domain as
\begin{equation}
    \rec_{j}^{i}[\n]= \sum_{\tau}\mathbf{s}^{i}_{\text{a}}[\n-\tau]\mathbf{g}^{r,z}_j[\tau] + \qv^{i}[\n],
    \label{eqn:rfn2}
\end{equation}
 We used square brackets, such as $[\n]$, to index vectors. Here, discretized receiver function $\rec_{j}^{i} \in \mathbb{R}^{\Nn_t}$, where $\Nn_t$ represents the total number of time samples. We normalized all the RFs so that the amplitude of the P wave at zero time lag is 1. Note that assuming that crustal effects are coherent across earthquakes in a bin, the earthquake index $i$ was not applied to $\mathbf{g}^{r,z}_j$.

\subsection{Multiple stations and datapoints for training }
Our network is trained using datapoints from multiple stations simultaneously. Thus, the total number of datapoints is the aggregate of all bins across all the stations.
Our main objective is to detect subtle variations in crustal effects between different datapoints, while minimizing the nuisance effects. The training dataset is the collection of all radial and transverse datapoints:
\begin{equation}
\label{eqn:traindata}
    \mathbf{T} = \{\rec_{j}\,\mid 1 \leq j \leq \Nn_b\} \cup \{\mathbf{t}_{j}\,\mid 1 \leq j \leq \Nn_b\},
\end{equation}
where $\rec_j$ and $\mathbf{t}_j$ denotes the radial and transverse datapoint associated with the $j$th bin, respectively. $N_b$ indicates the total number of bins of across all stations.
Each radial RF datapoint in the training set is 
given by the vector
\begin{equation}
\label{eq:datapoint}
 \rec_j= [\rec_j^{1}, \rec_j^{2}, \ldots, \rec_j^{\Nn_j}], 
\end{equation}
where we collect all RFs in a specific backazimuth epicentral distance bin associated with a particular station. $\Nn_j$ denotes the number of RFs in the $j$th bin.
The $\mathbf{r}_j$ is reshaped into a matrix with dimensions $\Nn_t \times \Nn_j$ for training input. 
Note that crustal effects are coherent across all RFs in a data point $\rec_j$.
%
Moreover, this construction of datapoints can be extended to RFs resulting from different seismic phases, such as S, SKS, and PKP, which is reserved for future study.

\section{Symmetric Variational  Autoencoder}
\label{sec:sym}
\begin{figure}
    \centering
    \includegraphics[scale=0.45,trim={0 0 3.7cm 0},clip]{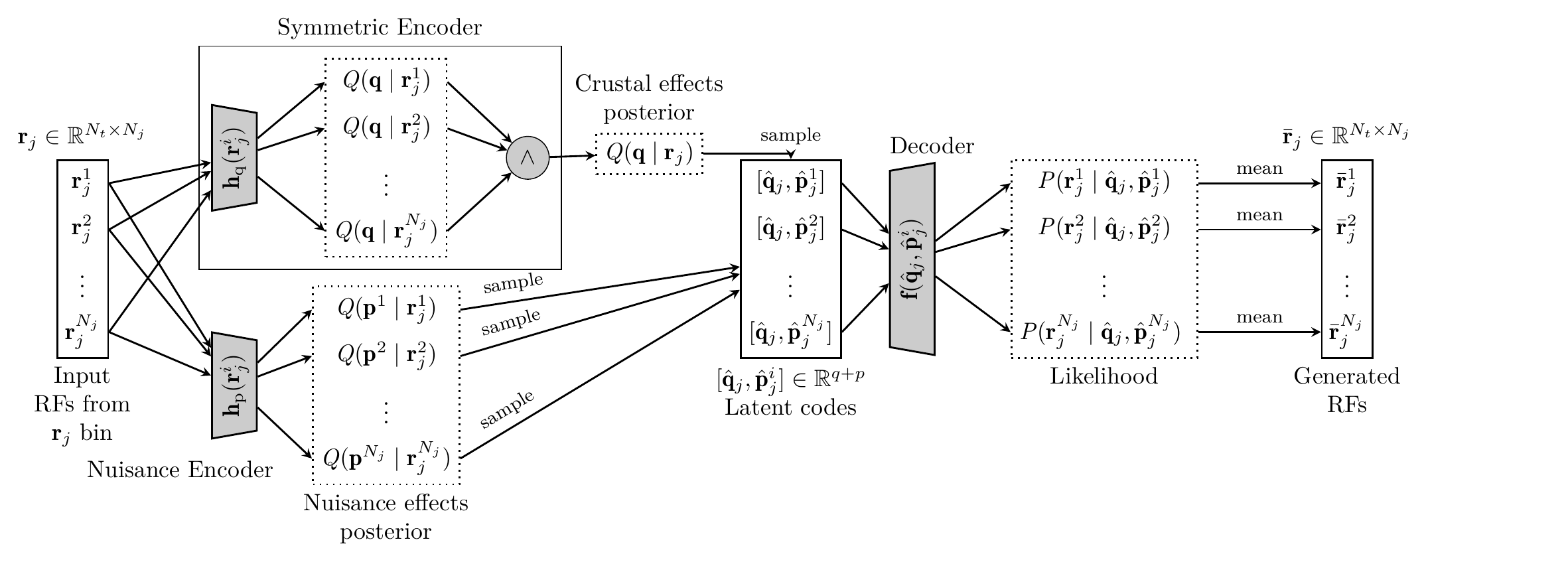}
    \caption{Schematic illustration of symmetric  variational  autoencoders (SymVAE). The input to the network takes is a group of RFs, i.e., a datapoint denoted as $\rec_j$, from a designated backazimuth-epicentral-distance bin. The symmetric encoder and the nuisance encoder output 
    posterior distributions associated with 
    the coherent crustal effects and nuisance effects, respectively. The decoder performs forward modeling to generate RFs using the sampled latent codes from the posterior. 
    }
    \label{fig:network}
\end{figure}

We will now focus on training the symmetric variational autoencoder (SymVAE) \citep{bharadwaj2024vsymae} using the datapoints constructed in the previous section. Appendix \ref{sec:vae} provides a detailed review of standard variational autoencoders~\citep{Kingma2014,bishop2023deep,prince2023understanding} --- 
we will expand on that discussion here.
The main advantage of SymVAE over the standard VAE lies in its ability to
learn a disentangled latent representation of the datapoints.
From the Appendix.~\ref{sec:vae},recall that the encoder maps each datapoint to the latent space.
The encoder of SymVAE is constrained to ensure that a few dimensions of latent space uniquely capture the coherent crustal effects,
while other dimensions capture the nuisance effects (see Fig.~\ref{fig:network}). In other words, the latent space is partitioned into the following two components.
\begin{description}
 
    \item[Crustal effects component.] This component, denoted using a vector of random variables $\coh$, captures the crustal effects on the receiver side. For a datapoint $\rec_j$ of $\mathbf{T}$, the RFs have a similar backazimuth and epicentral distance. These RFs are associated with raypaths that pass through the same receiver-side crust. This component represents the coherent crustal effects across all the RFs. As shown in Fig.~\ref{fig:network},
    the posterior associated with this
component, i.e. $\post(\coh\mid \rec_j)$, is output by the symmetric encoder network $\mathbf{h}_q$.

    \item[Nuisance effects component.] This component is RF specific, 
    as it represents the nuisance effects, due to the deconvolution process and seismic noise, unique to the specified RF.
    It is denoted 
    using a vector of random variables $\nuis^i$ corresponding to the
$i$th RF in $\rec_j$. As shown in Fig.~\ref{fig:network}, the posterior associated with $\nuis^i$, i.e., $\post(\nuis^i\mid\rec^i_j)$, is output by the nuisance encoder network $\mathbf{h}_p$.
\end{description}
When we say encoders output posterior distributions, it means they provide the parameters of these distributions, specifically a mean vector and diagonal covariance matrix for multivariate Gaussian distributions.
As an example, consider the nuisance encoder that is responsible for capturing the nuisance effects associated with each RF $\rec^i_j$.
The convolutional neural network $\mathbf{h}_p$, detailed in Appendix~\ref{sec:training}, determines the mean and diagonal covariance of the posterior distribution $\post(\nuis^i\mid\rec^i_j)$.

\subsection{Accumulation of crustal information using symmetric encoder}
The primary attribute of the SymVAE symmetric encoder is its ability to accumulate crustal information from multiple available RFs.
In other words, 
for a given datapoint $\rec_j$, each RF $\rec^i_j$ is considered an independent observation that contains information on crustal effects.
The symmetric encoder first generates a posterior distribution of crustal information for each RF within $\rec_j$ --- 
we denote the distribution associated with $i$th RF by $\post(\coh\mid\rec^i_j)$. The convolutional neural network $\mathbf{h}_q$ (detailed in Appendix~\ref{sec:training}) determines the mean and diagonal covariance of the posterior $\post(\coh\mid\rec^i_j)$.
The encoder then combines all these posterior distributions using the principle of \emph{conjunction of information states} as indicated by ~\citep{Tarantola2005}, where the  accumulated crustal information is represented through the posterior
\begin{equation}
    \post(\coh\mid \rec_j) = \post(\coh\mid \rec^{1}_j)\, \wedge \, \post(\coh\mid \rec^{2}_j)\, \wedge \cdots \varpropto \prod_{i} \, \post(\coh\mid \rec^{i}_j).
\end{equation}
Here, $\wedge$ denotes the conjunction, 
which means that as more independent information on crustal effects is collected through additional RFs within a datapoint, the precision of the crustal effects is enhanced. The conjunction limits the capture of nuisance effects, specific to individual RFs, aiding in the disentanglement of crustal effects.
Note that the accumulated information remains unchanged regardless of the order of the RFs within a datapoint, meaning that it is symmetric with respect to their ordering. 
As SymVAE models $\post(\coh\mid\rec^i_j)$ for each $i$ as a multivariate Gaussian with diagonal covariance, $\post(\coh\mid \rec_j,\theta_\qln)$ is also a multivariate Gaussian.

\subsection{Generating enhanced virtual receiver functions}
The generation of RFs from the latent space is performed by a convolutional decoder network $\mathbf{f}$ (details in Appendix~\ref{sec:training}). The decoder needs a latent code as input, which is constructed by drawing samples from the posterior distributions.
As shown in the Fig~\ref{fig:network},
the coherent code sampled from the posterior $\post(\coh\mid\rec_j)$ is labeled $\hat{\coh}_j$, and the nuisance code sampled from $\post(\nuis^i\mid\rec^i_j)$ is labeled as $\hat{\nuis}^i_j$.
The RF generated by the decoder will contain features associated with these latent codes
--- the coherent code sampled from $\post(\coh\mid\rec_j)$ represents the crustal effects of the datapoint $\rec_j$, while the nuisance code from $\post(\nuis^i\mid\rec^i_j)$
represents unique nuisance effects linked to the RF $\rec^i_j$. 
%
%
The decoder $\mathbf{f}$ provides the mean of the likelihood distribution $\likl(\rec^i_j\mid\hat{\coh}_j,\hat{\nuis}^i_j)$, a multivariate Gaussian with constant variance for all variables.
The mean of the likelihood distribution $\likl(\rec^i_j\mid\hat{\coh}_j,\hat{\nuis}^i_j)$ corresponds to the generated RF linked to the latent codes $\hat{\coh}_j$ and $\hat{\nuis}^i_j$.
As described in Appendix~\ref{sec:vae}, optimizing encoder and decoder parameters to maximize the evidence lower bound will not only ensure accurate reproduction of the RFs in the training dataset $\mathbf{T}$

\begin{figure}
    \centering
    \includegraphics{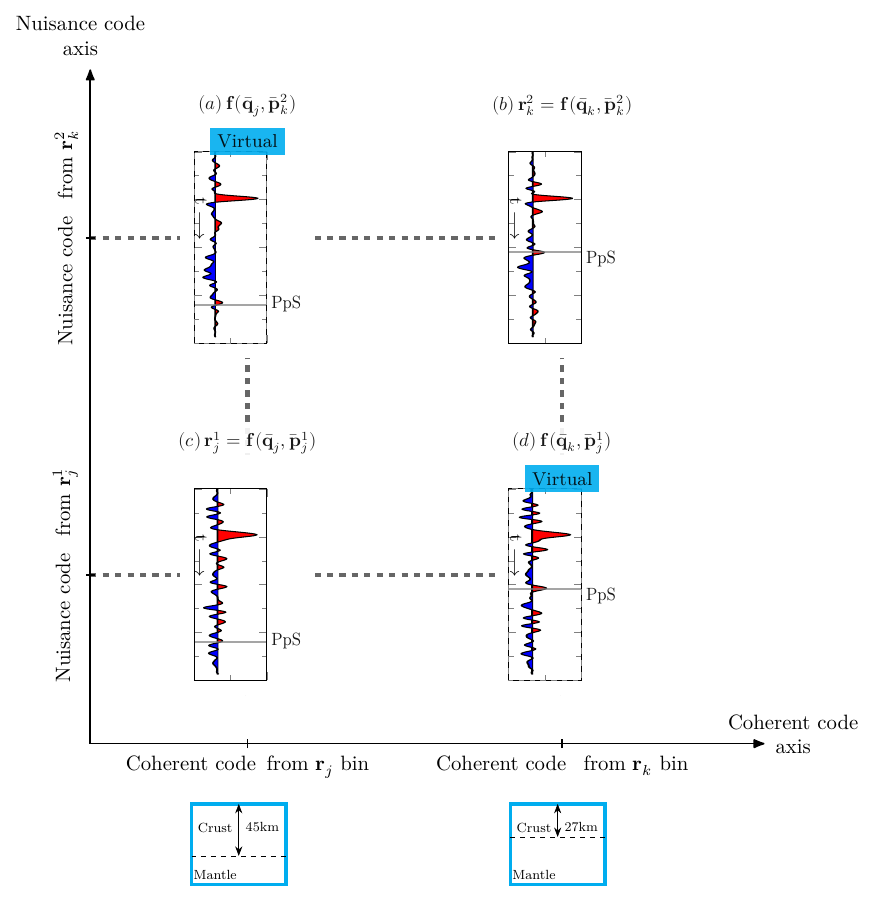}
    \caption{This figure illustrates the generation of new synthetic radial RFs using  hybrid latent code. The coherent code captures crustal effects, while the nuisance code captures the RF-specific nuisance effects. (b) shows a reconstructed radial RF, generated using coherent and nuisance code from bin $\rec_k$, associated with a station having a $27$ km thick crust. Similarly, (c) corresponds to bin $\rec_j$ and is related to a station with a $45$ km thick crust. (a) and (d) are virtual RFs generated by using coherent code from the other bin. The pairs (a, b) and (c, d) are observed to have the same nuisance effects but differ in crustal effects, as indicated by the delay time of backscattered S-wave (PpS). This is because they are generated using the same nuisance code but have different coherent codes.}
    \label{fig:redat}
\end{figure}

After training SymVAE, the code pertaining to the coherent crustal effects of the datapoint $\rec_j$ is given by the mean of the accumulated posterior $\post(\coh\mid\rec_j)$, which is denoted as $\mcoh_j$. The code related to nuisance effects for a specific RF $\rec^i_j$ is given by the mean of the posterior $\post(\nuis^i\mid\rec^i_j)$, denoted as $\mnuis^i_j$. We can create a hybrid latent code by combining coherent and nuisance from different datapoints. The decoder then uses this hybrid latent code to generate new RFs that correspond to the combined features indicated by the hybrid code. An example of generating new RFs is shown in Fig.~\ref{fig:redat}. In the figure, we considered two datapoints, $\rec_j$ and $\rec_k$. The RFs in $\rec_j$ are computed from a synthetic station with a crustal thickness of $45$ km, while the RFs in $\rec_k$ are computed from a synthetic station with a crustal thickness of $27$ km. More details on synthetic RFs is provided below. Fig~\ref{fig:redat}c illustrates the reconstructed RF $\rec^1_j=\mathbf{f}(\mcoh_j,\mnuis^1_j)$, generated by using the coherent code $\mcoh_j$ and the nuisance code $\mnuis^i_j$ from the same datapoint $\rec_j$. Similarly, in Fig~\ref{fig:redat}b, the reconstruction is from datapoint $\rec_k$. These reconstructed RFs are the original RFs in the training data. On the other hand, the new RF in Fig~\ref{fig:redat}a, represented as $\mathbf{f}(\mcoh_j,\mnuis^2_k)$, is generated using the coherent code $\mcoh_j$ from datapoint $\rec_j$ and the nuisance code $\mnuis^2_k$ from RF $\rec^2_k$ of datapoint $\rec_k$. Similarly, the new RF in subplot Fig~\ref{fig:redat}d, represented as $\mathbf{f}(\mcoh_k,\mnuis^1_j)$, combines the coherent code $\mcoh_k$ from $\rec_k$ with the nuisance code $\mnuis^1_j$ from RF $\rec^1_j$ of datapoint $\rec_j$.
Notice that the delay time of the backscattered S-wave (PpS), which depends on the thickness of the crust, corresponds to the coherent code. For example, the new RF in Fig~\ref{fig:redat}a and reconstructed RF in Fig~\ref{fig:redat}b exhibit different delay times for PpS while sharing the same nuisance effects. This is because they are generated using the same nuisance code from RF $\rec^2_k$, but different coherent codes. Similarly, the new RF in Fig~\ref{fig:redat}d and reconstructed RF in Fig~\ref{fig:redat}c share the same nuisance effects but different delay times for PpS. In addition, reconstructed RF $\rec^1_j$ in Fig~\ref{fig:redat}b and new RF $\mathbf{f}(\mcoh,\mnuis^1_j)$ in Fig~\ref{fig:redat}d share the same crustal effects i.e. delay time of PpS but different nuisance effects, as they are generated using the same coherent code $\mcoh_k$ from datapoint $\rec_k$.

We have seen that RFs sharing the same coherent crustal effects but exhibiting different nuisance effects can be generated in SymVAE. Note that the RF in Fig~\ref{fig:redat}b exhibits less nuisance effects compared to the RF in Fig~\ref{fig:redat}d. This indicates that certain new RFs have minimal nuisance effects allowing clarity in observing the crustal effects. Similarly to the previous example, we can generate a collection of RFs using identical coherent code from data point, say $\rec_j$ but different nuisance code sampled from the nuisance posterior distribution $\post(\nuis^i\mid\rec^i_j)$. The sampled nuisance code, denoted by $\Ncode$, may not exactly correspond to the nuisance effects of RFs from the training data, but it still represents realistic nuisance effects. For each sampled nuisance code $\Ncode$, the generated new RF can be represented as
\begin{equation}
    \vir^{\Ncode}_j =\mathbf{f}(\mcoh_j,\Ncode),
\end{equation}
where $\mcoh_j$ is the coherent code from datapoint $\rec_j$. These new RFs exhibit different variety of nuisance effects, but same coherent crustal effects from the datapoint $\rec_j$. Our goal now is to identify which ones can better serve the purpose of extracting the crustal effects from the datapoint $\rec_j$. Specifically, we want to obtain new RF with minimal nuisance effects related to data point $\rec_j$. This enhanced new RF is called virtual RF. To obtain virtual RF, we use the Kullback-Leibler (KL) divergence,
\begin{equation}
    H (\vir^{\Ncode}_j) = D_{KL} \left( \post(\coh\mid\rec_j)\, || \,\post(\coh\mid\vir^{\Ncode}_j)\right).
    \label{eqn:dkl}
\end{equation}
This serve as a tool for finding new RFs that predominantly capture crustal effects with minimal nuisance effects~\citep{bharadwaj2024vsymae}. Here, $\post(\coh\mid\vir^{\Ncode}_j)$ denotes the crustal information encapsulated in the new RF $\vir^{\Ncode}_j$, while $\post(\coh\mid\rec_j)$ refers to the accumulated crustal information derived from all RFs in the datapoint $\rec_j$. The KL divergence evaluates the degree of discrepancy between two probability distributions, effectively quantifying the extent of crustal effect information that is lost when the actual RFs in $\rec_j$ are replaced by the new RF $\vir^{\Ncode}_j$. Consequently, a lower KL divergence $H(\vir^{\Ncode}_j)$ indicates a superior preservation of the accumulated crustal effects in $\vir^{\Ncode}_j$. We estimate the best nuisance code $\Ncode$  by optimizing the KL divergence $H (\vir^{\Ncode}_j)$ w.r.t. $\Ncode$. We select the RF's nuisance code that results in the least KL divergence as the starting point for $\Ncode$. An optimization process then iteratively updates $\Ncode$ until convergence, producing a new RF with minimal nuisance effects for that datapoint. This is similar to guiding the decoder network to produce a new RF with crustal effects close to the accumulated crustal effects of the datapoint. This enhanced virtual RF related to datapoint $\rec_j$ can be expressed as 
 \begin{equation}
     \vir^{\text{virt}}_j=\mathbf{f}(\mcoh_j,\Ncode^{\text{virt}}),
 \end{equation}
 where $\Ncode^{\text{virt}}$ denotes the optimal nuisance code and $\mcoh_j$ denotes the coherent code from datapoint $\rec_j$. Until now we considered one datapoint and obtained the corresponding optimal nuisance code. This optimal nuisance code can be reused for other datapoints of the same station. In practice, we select the datapoint with the highest RF count among all datapoints in a station to determine the optimal nuisance code. This optimal nuisance code is then used to generate virtual RFs for the other datapoints within the station. For example, virtual RFs for all the datapoints belonging to station $A$ is given by a set
\begin{equation}
 \label{eqn:redatuming}
     \mathbf{Y}_A=\{\vir^{\text{virt}}_j \mid j \;\text{corresponds to station} \;A\}= \{\mathbf{f}(\mcoh_j,\Ncode^{\text{virt}})\mid j \;\text{corresponds to station}\;A\}
 \end{equation}
where $\Ncode^{\text{virt}}$ denotes the optimal nuisance code derived from the datapoint with the highest number of RFs in station $A$. Note that the virtual RFs are generated using different coherent codes specific to the datapoints but the same optimal nuisance code. Similarly, we can obtain the optimal nuisance code $\Ncode$ for every station and generate virtual RFs for all stations:
\begin{eqnarray}
 \label{eqn:virtualRF}
 \nonumber
     \mathbf{Y} &=& \bigcup_{\kappa\in\, \text{stations}} \,\{\vir^{\text{virt}}_j \mid j \;\text{corresponds to station} \;\kappa\} \\
     &=& \bigcup_{\kappa\in\, \text{stations}} \,\{ \mathbf{Y}_\kappa\}
 \end{eqnarray}   
Keep in mind that the generation of virtual RFs for data points at each station is independent, without focusing on lateral coherence between stations. In the following sections, we assess the quality of these virtual RFs compared to those obtained by averaging techniques using synthetic and real seismic data.

\section{Synthetic Experiment}
\label{sec:synex}

In this section, we evaluate our method using synthetic RFs,
which contain realistic nuisance effects. 
These experiments guided our selection of network architecture and training hyperparameters, detailed in the Appendix~\ref{sec:training}. 
We used a velocity model, as detailed in Tab.~\ref{tab:model1}, which features a single crustal layer with $10\%$ transverse anisotropy, above an isotropic mantle. 
The anisotropic crustal layer has the fast axis oriented at an azimuth angle of $0^{\circ}$N. Six stations (Stations 1--6) were considered in this experiment, with the thickness of the crustal layer progressively increasing from Station 1 to Station 6. The anisotropy leads to systematic variations of the converted waves with respect to the backazimuth. 
The creation of synthetic RFs involves several steps detailed here.
\begin{description}
\item \textbf{Crustal impulse responses}. At each station, the crustal impulse responses for the vertical, radial, and transverse components were calculated using the PyRaysum software~\citep{Bloch_Audet_2023}. We analyze $10^{\circ}$-interval bins for backazimuth and epicentral distance. 
The backazimuth spans from $0^{\circ}$ to ${360}^{\circ}$, and the epicentral distance spans from $50^{\circ}$ to $60^{\circ}$.
Within each bin, we generated crustal impulse responses using randomly chosen values of the backazimuth and epicentral distance within the defined range. 
The number of impulse responses assigned to each bin was randomly determined, ranging from 5 to 35. That means that certain bins experience earthquakes more frequently than others.


\item \textbf{Earthquake signatures}. To create synthetic seismograms, the components of each crustal impulse response were convolved with a unique signature of the earthquake source obtained from real seismograms, after trimming a 40\,s window centered around the P-wave arrival. This approach not only mimics the bandwidth variations and zeros found in the source spectrum typical of real earthquakes but also confirms
that there are no repeated earthquake occurrences.


\item \textbf{Adding ambient seismic noise}. Ambient noise from seismic stations in the US was added to all three components independently. The noise level was adjusted to maintain a specified signal-to-noise ratio (SNR), which was randomly sampled from a truncated log-normal distribution to the interval from $0.01$ to $2$.
We calculated the SNR by calculating the ratio of the root-mean-squared amplitude $10$\,s before and after the P-wave arrival on the vertical component seismogram.

\item \textbf{Deconvolution and datapoints}. Frequency domain deconvolution was employed to generate synthetic radial and transverse RFs.
The deconvolution process was regularized using a water level parameter of $0.01$ and a Gaussian filter with a width of $5$. 
Finally, RFs within each bin were grouped to create datapoints, as described in Eq.~\ref{eq:datapoint}. A total of $36\times6\times6$ synthetic datapoints, with all receivers included, were used to train the SymVAE network.


\end{description}

\begin{table}
\caption{Parameters of crustal velocity model, which includes a layer with $10\%$ transverse anisotropy, with the fast-axis along $0^{\circ}$N.}
\begin{center}
    \begin{tabular}{|c|c|c|c|c|c|c|c|}
    \hline
         & Thickness (km) & Vp (km/s) & Vs (km/s) & ${\rho}$ (kg/m3) & Aniso. (${\%}$) & Trend (${^\circ}$) & Strike (${^\circ}$) \\
    \hline\hline 
    Layer 1 & 27.0 & 6.0 & 3.45 & 2.8 & 10.0 & 0 & 0 \\ 
    Layer 2 & NA & 8.04 & 4.47 & 3.6 & 0 & 0 & 0 \\  
    \hline
    \end{tabular}

    \label{tab:model1}
    \end{center}
    \end{table}

\begin{figure}
    \centering
    \includegraphics[trim={0cm 0.5cm 0 0.2cm},width=\textwidth]{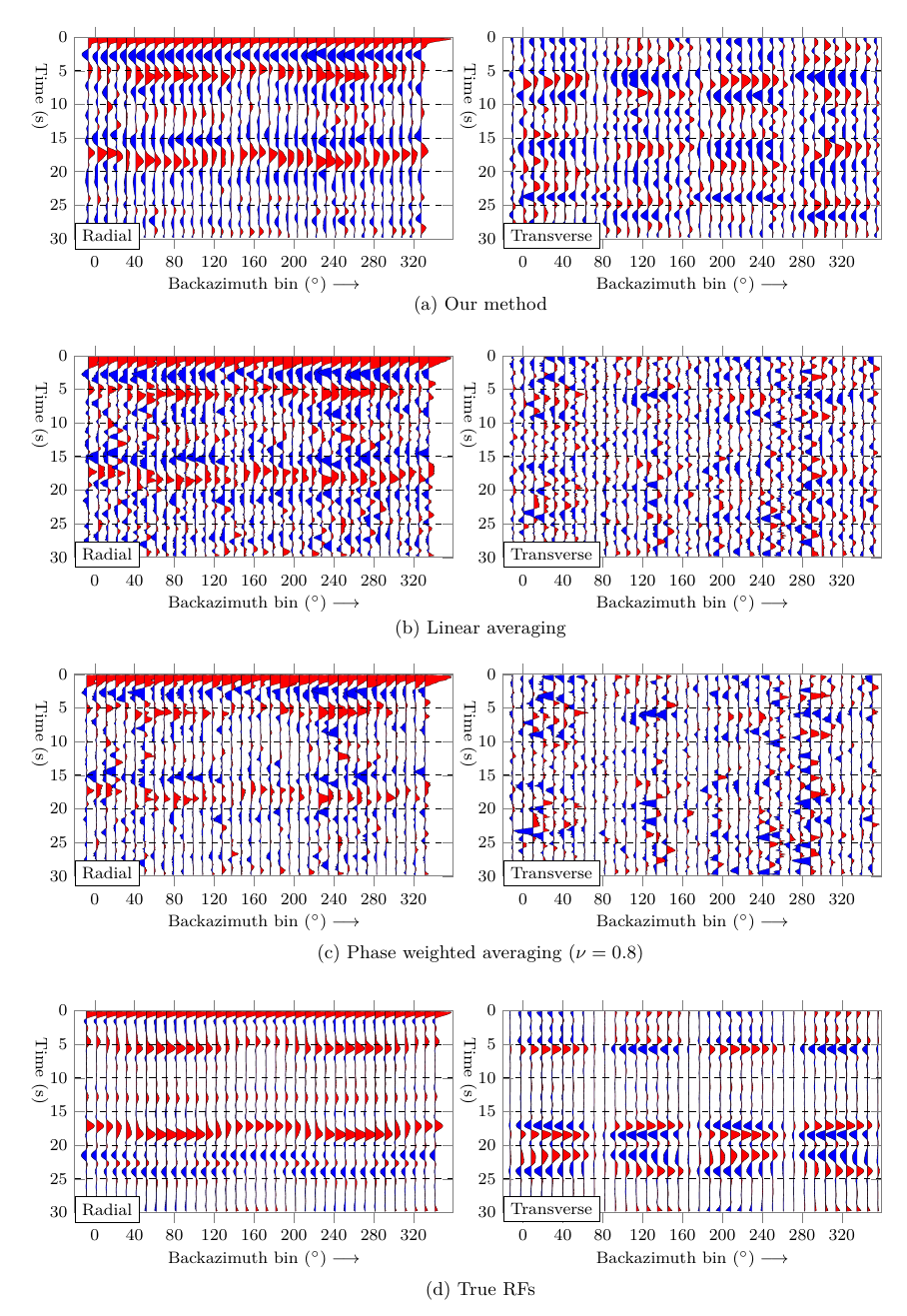}
    \caption{Radial and transverse receiver functions from the synthetic experiment, derived using (a) SymVAE (b) linear bin-wise averaging and 
    (c) phase-weighted averaging, are compared to (d) true responses.
    Only Station 4 with a $40.5$km thick crustal layer is considered.
    }
\label{fig:model1}
\end{figure}

Following network training, we generate the virtual RF for each station's backazimuth bin according to Eq.~\ref{eqn:redatuming} --- Fig.~\ref{fig:model1}a shows these RFs for both radial and transverse components.
%
To assess the accuracy of the synthetic RFs, we used true RFs obtained with the same steps as above but employed a Gaussian source wavelet and omitted noise.
The true RFs are plotted in Fig.~\ref{fig:model1}d.
The virtual RFs generated by the network exhibit reduced nuisance effects compared
to linear and phase-weighted averaging (power $\nu$ = $0.8$), plotted in Figs.~\ref{fig:model1}b and \ref{fig:model1}c, respectively.
This is clear because the normalized correlation coefficient (NCC) $\ncc$ between the virtual RFs and true RFs exceeds that of linear or phase-weighted averaging, as shown in Fig.~\ref{fig:mse1} for Station 4 (see Appendix.~\ref{sec:syn-eval} for more detail on NCC).   In this plot, note that the NCC value remains largely unaffected by the number of earthquakes (or RFs) in a bin, even when averaging is ineffective.
We excluded the P arrival while computing the NCC.
Finally, notice that the transverse virtual RFs are significantly more enhanced than radial RFs.
The results are uniform across all receivers, with detailed findings available in the supplementary material.
%

\begin{figure}
\centering
\includegraphics[width=\textwidth]{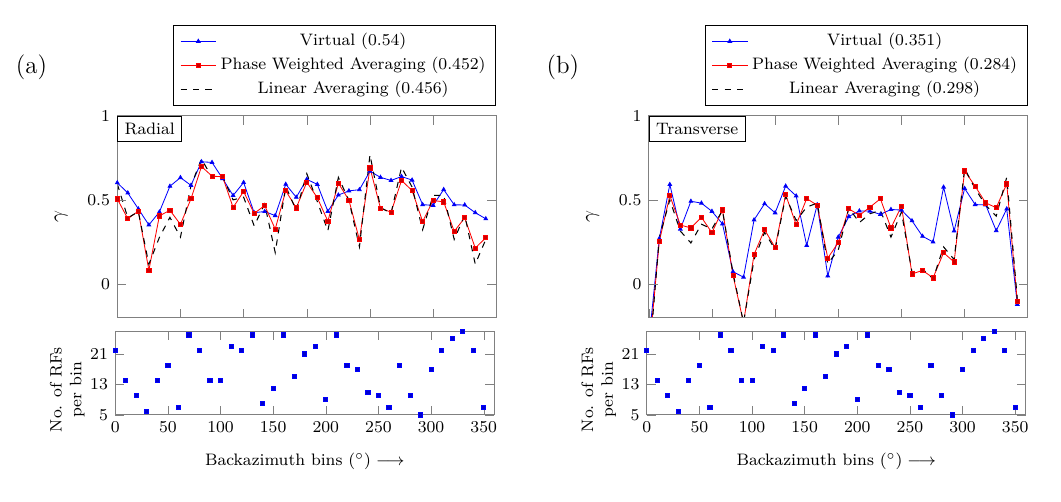}
\caption{The quality of the (a) radial and (b) transverse RFs, plotted in Fig.~\ref{fig:model1}, is assessed using the normalized correlation coefficient $\ncc$ ----detailed in Apppendix.~\ref{sec:pcc}. 
On average, the coefficient $\ncc$ of the SymVAE-generated virtual RFs is higher compared to linearly averaged and phase-weighted averaged RFs, indicating higher correlation with the true RFs.
\label{fig:mse1}}
\end{figure}



\section{Application on real data}
We evaluated our method using teleseismic data recorded in two geologically distinct settings: 1) the Cascadia Subduction Zone, where the Pacific plate is gradually subducting beneath the North American plate, and 2) Southern California, where the crustal thickness is known to be heterogeneous across major strike-slip faults.
For a given RF, derived using SymVAE or bin-wise averaging,
we assessed its quality by computing the mean normalized correlation coefficient (MNCC) $\ncc_{\text{m}}$ w.r.t. RFs within datapoint --- see Appendix~\ref{sec:real-eval} for more details.


Our results demonstrate that our method is scalable across various datasets and regions, consistently maintaining good performance. 

\subsection{Data preparation}
 To implement our approach in real data, teleseismic earthquakes recorded by $42$ seismic stations in southern Vancouver Island, British Columbia, and the Olympic Peninsula, belonging to the POLARIS~\citep{POL}, C8~\citep{c8}, CN~\citep{cn} and UW~\citep{uw} networks were used. These seismic stations, shown in Fig.~\ref{fig:casmap}, cover the Vancouver Island segment of the Cascadia Subduction Zone. We also used teleseismic data from $243$ seismic stations  located in Southern California ~\citep{Caltech1926} belonging to CI network ~\citep{Caltech1926} (see Fig.~\ref{fig:calimap}). 
  Earthquakes with epicentral distances ranging from $30^{\circ}$ to $100^{\circ}$ and a minimum moment magnitude of $5.5$ were used. The seismograms were rotated to radial and transverse components, then bandpass filtered using $0.03$ -- $3.0$ Hz, and trimmed at $20$\,s before the PREM-calculated P arrival time and  $50$\,s after.
Radial and transverse RFs were generated for all available seismograms, regardless of the SNR, using water-level deconvolution with a Gaussian factor of $5.0$ and a water level of $0.01$. In each station, the calculated RFs were binned according to the backazimuth angle and epicentral distance. Backazimuth angle interval of $8^{\circ}$ and epicentral distance interval of $5^{\circ}$ were used. Excluding bins with fewer than $2$ RFs resulted in $20,115$ bins (datapoints) across $285$ stations.

The training process for these datapoints is elaborated in Appendix~\ref{sec:training}.
After training, optimized virtual RFs for each datapoint or bin were generated station-wise according to Eq.~\ref{eqn:redatuming}. The subsequent subsections discuss the geological complexity of the regions and evaluate the effectiveness of SymVAE-generated RFs in delineating the structures compared to those averaged linearly.

\subsection{Cascadia subduction zone}
RF studies have been widely utilized to characterize crustal structures in subduction zones~\citep{Cassidy1995,Bloch2023,Savage2007,ShiomiK}.
However, the complex structural settings in the subduction zones, which include low-velocity zones, partial melting, and accretionary prisms, present significant obstacles in the extraction of crustal and upper mantle structures from RFs. 
Supervised denoising methods do not generalize well to denoise RFs from these subduction zones due to the unique structures and complexities of each zone, which differ from the training models. The averaging of RFs assumes that the nuisance effects follow a Gaussian distribution, making it ineffective when the number of RFs per datapoint is less. This is particularly important for temporary stations that record a limited number of earthquakes.
The Cascadia Subduction Zone, where the Juan de Fuca oceanic plate is subducted beneath the North American continental plate, has drawn significant research attention, similar to other subduction zones. This is due to its complex slab morphology, association with seismic hazards, and geodynamic processes \citep{Langston1979,Nicholson2005,Bloch2023}.
 Using receiver functions, 
~\cite{Bloch2023} and ~\cite{Audet2009Seismic} characterized the subducting oceanic slab as a system of two dipping layers: a low-velocity zone overlaying a higher-velocity zone.
Similar to conventional RF studies, these studies excluded noisy seismograms based on the signal-to-noise ratio (SNR) and linearly averaged the resulting RFs within a bin. 
 In contrast, our approach utilized all available RFs, irrespective of SNR, and generated virtual RFs for the extraction of converted phases from both radial and transverse components.

\begin{figure}
\centering
\includegraphics[width=\textwidth]{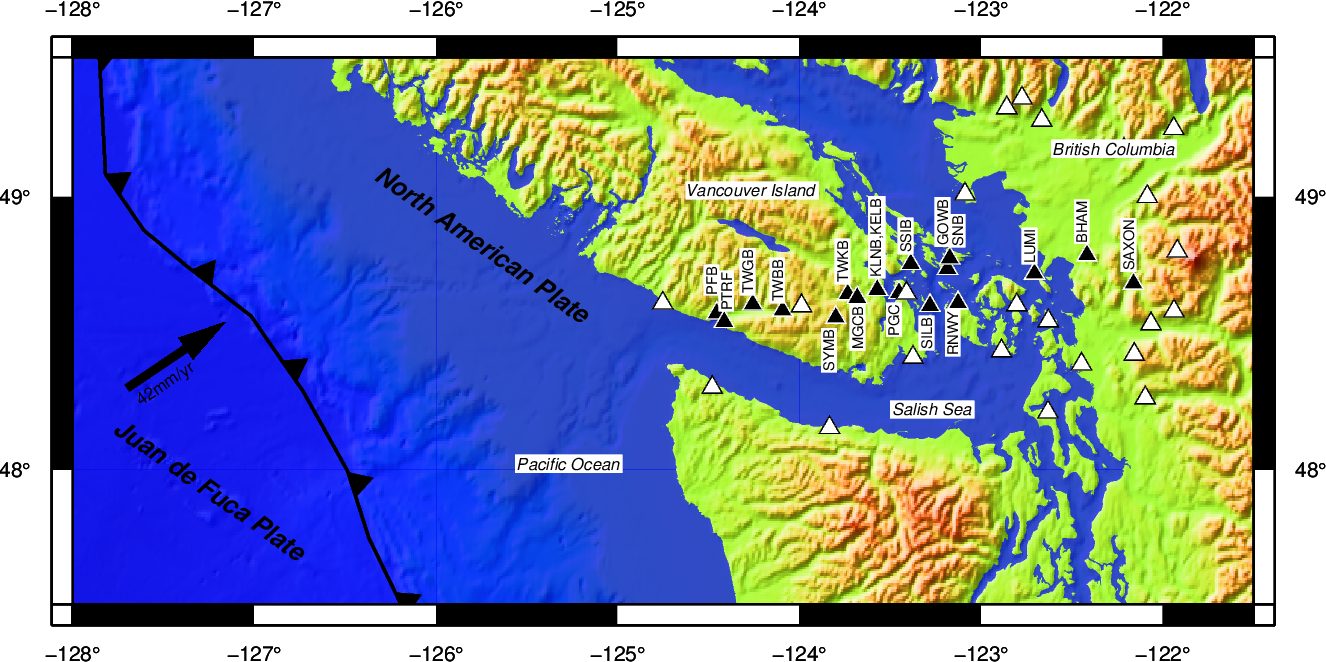}
\caption{Map illustrating the tectonic setting and positions of the stations (represented by black and white triangles) utilized in the study. This map also depicts the subduction of the Juan de Fuca plate beneath the North American plate, occurring at a rate of $42$ mm/yr ~\citep{DeMets1994} (indicated by a black arrow). The solid black line adorned with triangles marks the subduction boundary. Stations are from the POLARIS ~\citep{POL}, C8 ~\citep{c8}, CN ~\citep{cn}, and UW ~\citep{uw} networks. While all stations are utilized for training, this paper specifically plots the receivers marked by black triangles that are approximately positioned along a transect. 
}
\label{fig:casmap}
\end{figure}

\begin{figure}
\centering
\includegraphics[width=\textwidth]{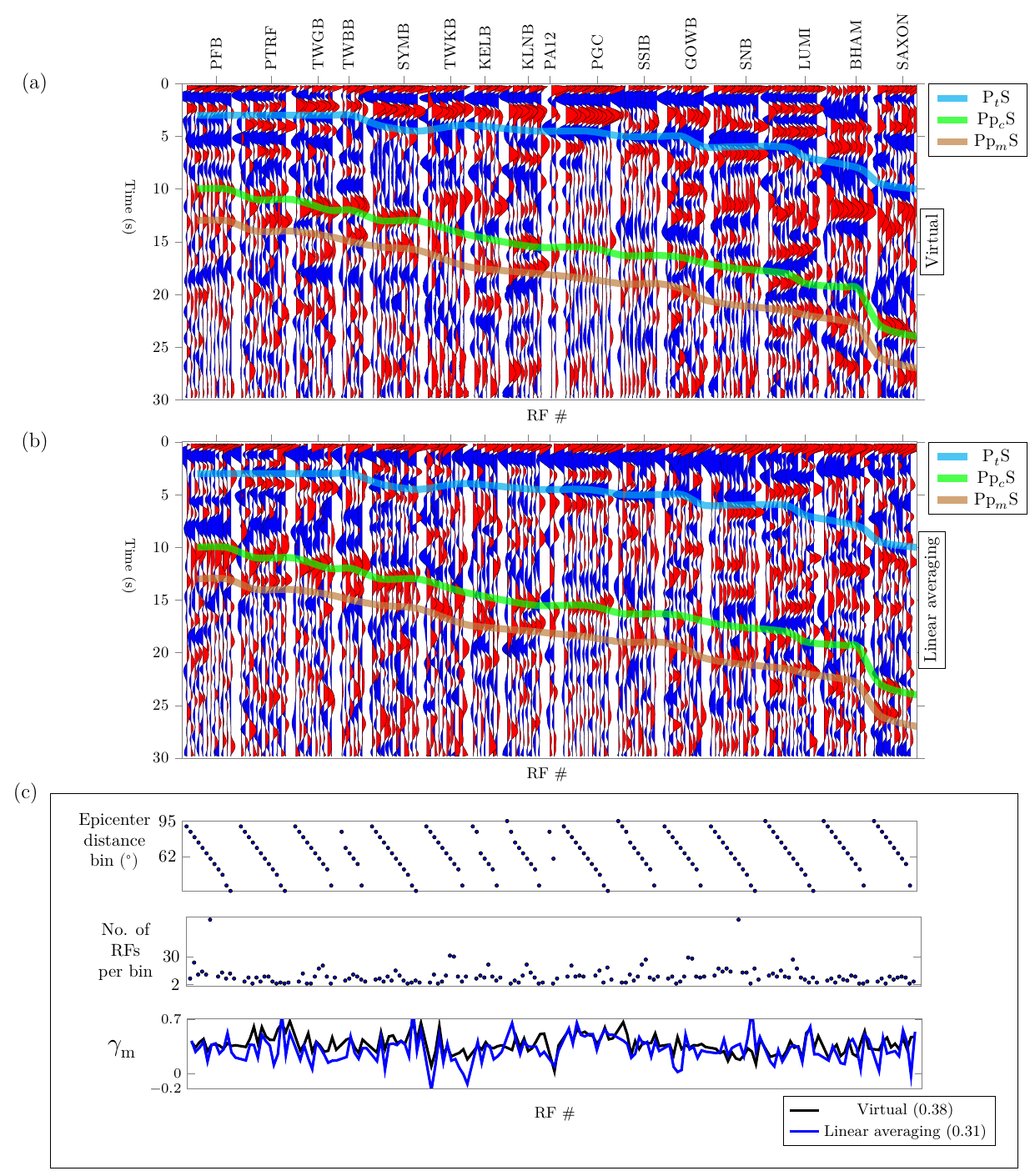}
\caption{(a) Virtual and (b) linearly averaged radial RFs at seismic stations (Fig.~\ref{fig:casmap}) tranversing the Cascadia subduction zone. (c) shows epicentral distance bin, number of RFs in the bin and  $\ncc_{\text{m}}$ (mean normalized correlation coefficient) corresponding to RFs in (a) and (b). Key seismic phases from the subducted slab  (highlighted by colored lines) are more clear and resolved in virtual RFs than linearly averaged RFs. The notation for these phases follows ~\cite{Bloch2023}, as given in Tab.~\ref{tab:phases_notation}. Virtual RFs generally show a higher coefficient $\ncc_{\text{m}}$ than averaged RFs, indicating increased performance of SymVAE.}
\label{fig:casrad}
\end{figure}

\begin{figure}
\centering
\includegraphics[width=\textwidth]{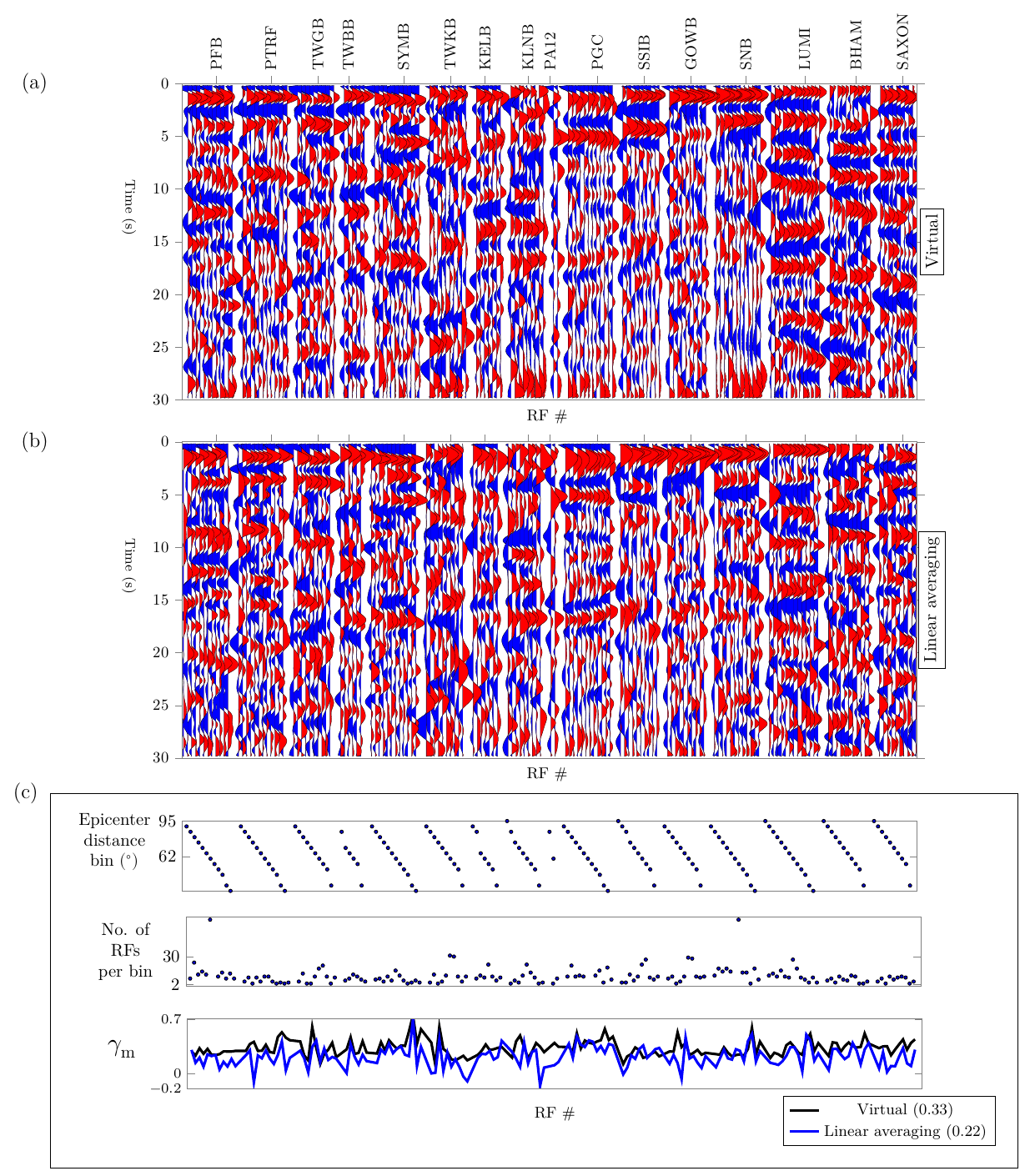}
\caption{Similar to Fig.~\ref{fig:casrad}, but pertaining to transverse RFs. It is important to observe that the virtual RFs exhibit greater complexity and enhancement, allowing for the converted phases to be smoothly traced along the transect. $\ncc_{\text{m}}$ (mean normalized correlation coefficient) of virtual RFs are higher compared to that of linearly averaged RFs, indicating its higher quality.}
\label{fig:castran}
\end{figure}

 To interpret and visualize the dip of the subduction slab, we have chosen 16 stations located along a transect in southern Vancouver Island, as shown in Fig.~\ref{fig:casmap}.
We also computed the averaged RF at each datapoint (or bin) for comparison with the generated virtual RF. Figs.~\ref{fig:casrad} and \ref{fig:castran} show station-wise plots of linearly averaged and virtual RFs for the profile in Fig.~\ref{fig:casmap}, organized by increasing epicentral distance bins. These RFs are from the same backazimuth bin $296^{\circ}$ to $304^{\circ}$. The figure also shows the mean normalized correlation coefficient (MNCC) $\ncc_\text{m}$ of virtual and averaged RFs (see Appendix~\ref{sec:pcc} for details on MNCC).  The virtual RFs generated are largely in alignment compared to ~\cite{Bloch2023} and ~\cite{Audet2009Seismic}, although they exhibit a greater coverage of the RFs. 
In addition, virtual RFs delineate the crustal structure in the subduction zone far better than linear-averaged RFs, as indicated by their higher MNCC. 
~\cite{Bloch2023} parameterized the subducting Juan de Fuca plate as a slab with three velocity interfaces: a negative velocity contrast at the top and two positive velocity contrasts below. In Fig.~\ref{fig:casrad}, the converted seismic phases reported by \cite{Bloch2023} (see Tab.~\ref{tab:phases_notation}) from the subducting slab (highlighted by colored lines) are exhibited clearly and consistently across the stations in the virtual RFs. 
 Notably, our approach was able to produced more number of RFs across varying backazimuths and epicentral distances, and effectively delineated the forward-scattered S-waves (between $0$ to $10$ s) more accurately than the RFs calculated by \cite{Bloch2023} (see supplementary material).
In all stations, Pp\textsubscript{c}S and Pp\textsubscript{m}S phases, which originated from the middle velocity contrast and bottom velocity contrast (Moho), are properly resolved in time in virtual RFs, unlike in averaged RFs. Regardless of how many RFs are in the bins (or datapoints), the virtual RFs maintain consistency across epicentral bins at each station. This consistency is not observed in averaged RFs, which are distorted by nuisance effects. Furthermore, virtual RFs show significant improvement in extracting crustal effects in transverse RFs, which is difficult to achieve by conventional RF technique because of the low SNR of the transverse component. Fig.~\ref{fig:castran} shows the virtual transverse RFs for the same datapoints in Fig.~\ref{fig:casrad}. These virtual RFs shows superior quality and consistency compared to average RFs, regardless of the number of RFs in the bins. This is indicated by higher MNCC $\ncc_\text{m}$ of virtual transverse RFs compared to linearly averaged RFs.   

\begin{figure}
\centering
\includegraphics[width=\textwidth]{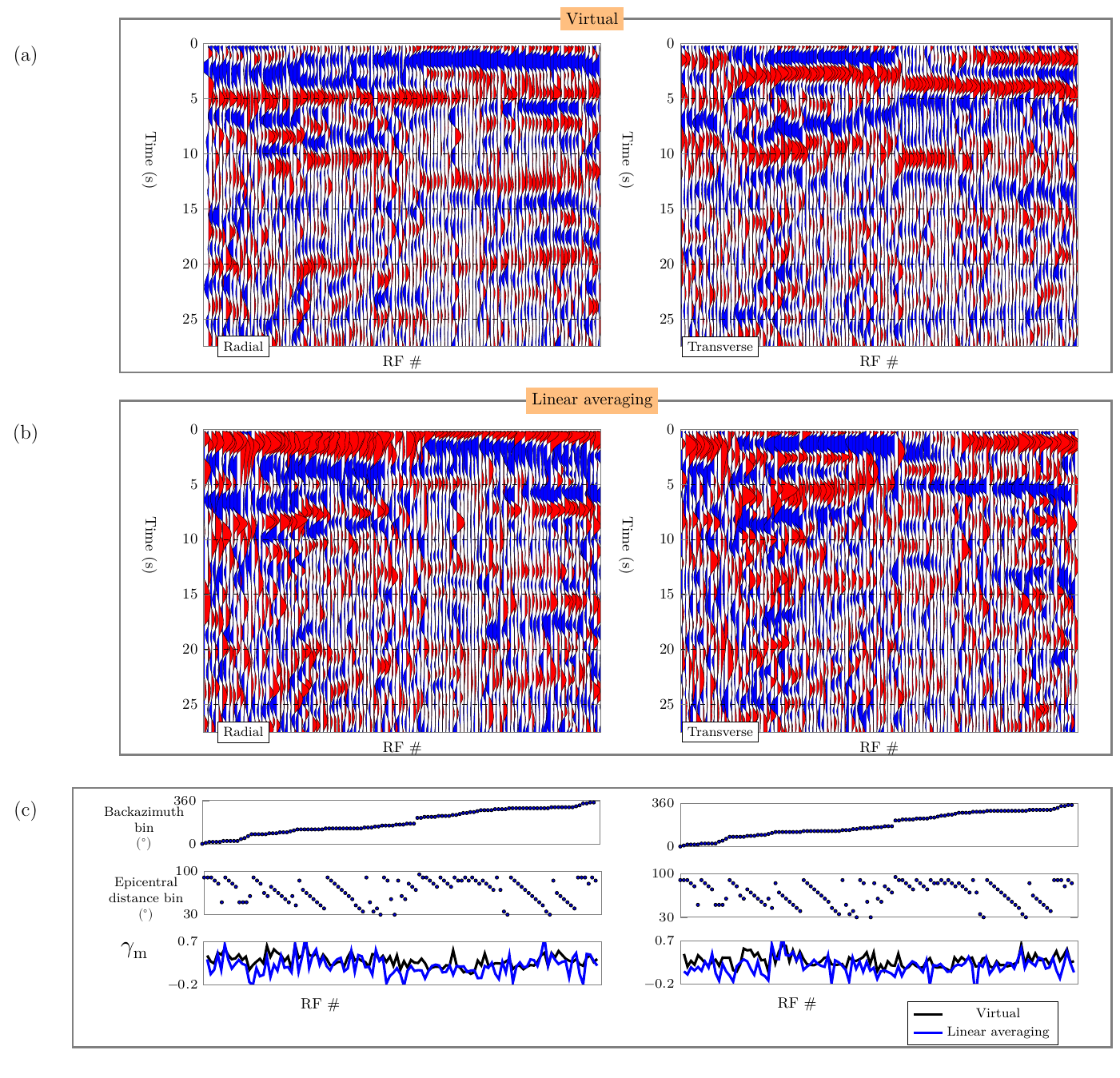}
\caption{(a) Virtual RFs from SymVAE and (b) RFs after linear averaging for RF bins from seismic station SNB (see Fig.~\ref{fig:calimap}), plotted by backazimuth angles. (c) shows the backazimuth angle bin, epicentral distance bin and $\ncc_\text{m}$  (mean normalized correlation coefficient)  corresponding to RFs in (a) and (b). Virtual RFs show clearer seismic arrivals compared to linearly averaged RFs and generally have higher $\ncc_\text{m}$.}
\label{fig:snb}
\end{figure}

\begin{figure}
\centering
\includegraphics[width=\textwidth]{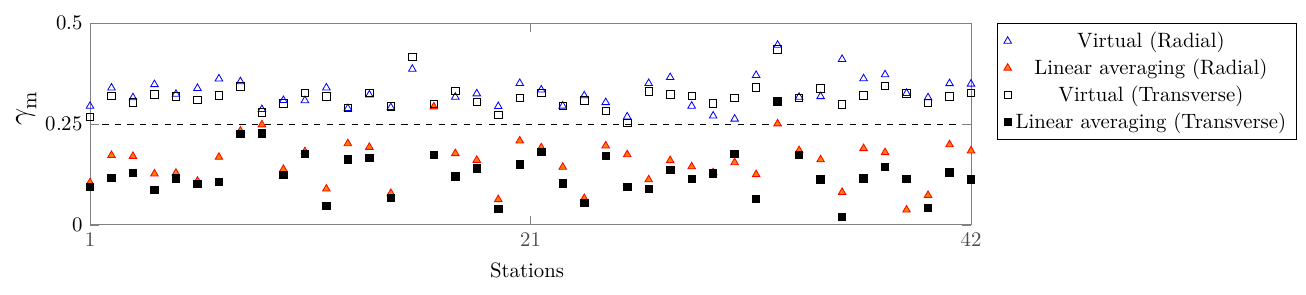}
\caption{For each station involved in the Cascadia case study, $\ncc_\text{m}$ (mean  normalized correlation coefficient) is averaged across all backazimuth and epicentral distance bins. Details of the correlation metric calculation are in Appendix~\ref{sec:pcc}. SymVAE-generated virtual RFs have higher $\ncc_\text{m}$ than linearly averaged RFs across the stations, highlighting the scalability and consistent performance of SymVAE.}
\label{fig:cascrr}
\end{figure}

Fig.~\ref{fig:snb} illustrates the virtual and averaged RFs for station SNB ordered with respect to the backazimuth angle. The virtual RFs correspond with the averaged RFs but provide more detail. Seismic signals are more pronounced in the virtual radial RFs, especially within the $0$ to $10$ s time window. Polarity reversal in the transverse RF are distinctly observed in virtual RFs compared to averaged RFs. The increased effectiveness of our method is signified by the generally higher MNCC $\ncc_\text{m}$ of the virtual RFs. Additionally, we computed the average MNCC across datapoints per station. The consistent higher MNCC $\ncc_\text{m}$ across stations, as depicted in Fig.~\ref{fig:cascrr}, validate the robustness of our method.


\begin{table}
  \centering
  \begin{tabular}{|c|c|}
    \hline
    Phase & Notation \\ \hline
    Forward scattered S wave from top velocity contrast & P\textsubscript{t}S \\ \hline
    Backward scattered S wave from middle velocity contrast & Pp\textsubscript{c}S \\ \hline
    Backward scattered S wave from bottom velocity contrast & Pp\textsubscript{m}S \\ \hline
  \end{tabular}
  \caption{This table illustrates the phases and their respective symbols as referenced in ~\cite{Bloch2023}. According to ~\cite{Bloch2023}, the subducting slab beneath the southern Vancouver Island is conceptualized with two distinct layers: one being a low-velocity layer resting above the oceanic crust. This configuration causes a negative velocity contrast at the top interface and two positive velocity contrasts: one occurs between the low-velocity layer and the oceanic crust, and the other occurs between the oceanic crust and the mantle.}
  \label{tab:phases_notation}
\end{table}

\subsection{Southern California}

Southern California has a complex network of active faults that mark the boundary between the Pacific and North American plates. The main fault lines include the San Andreas Fault (SAF), the San Jacinto Fault Zone (SJFZ), and the Elsinore Fault (EF), all of which run nearly parallel to the coast. The area is densely monitored by seismic stations \cite{Caltech1926}. Multiple studies have been conducted to explore the crustal structure and seismogenic processes \citep{hauksson2011crustal, SHAW20151, Wang2018}. RFs have been pivotal in defining the regional Moho geometry \citep{Zhu2000, zhu2000moho, Ozacar2009, Ozakin2015}. \cite{Ozakin2015} identifies significant undulations and sudden depth shifts of the Moho in the NE–SW direction along the plate boundary, showing vertical offsets of up to $\sim$8 km along the SAF and SJFZ. Previous RF analyses discarded noisy seismograms and RFs to increase data quality, reducing data coverage and affecting RF data reproducibility in the process. We applied our method to extensive teleseismic data from the dense network, creating virtual RFs without discarding seismograms or RFs, while maintaining both qualitative and quantitative data quality.

\begin{figure}
\centering
\includegraphics[width=0.7\textwidth]{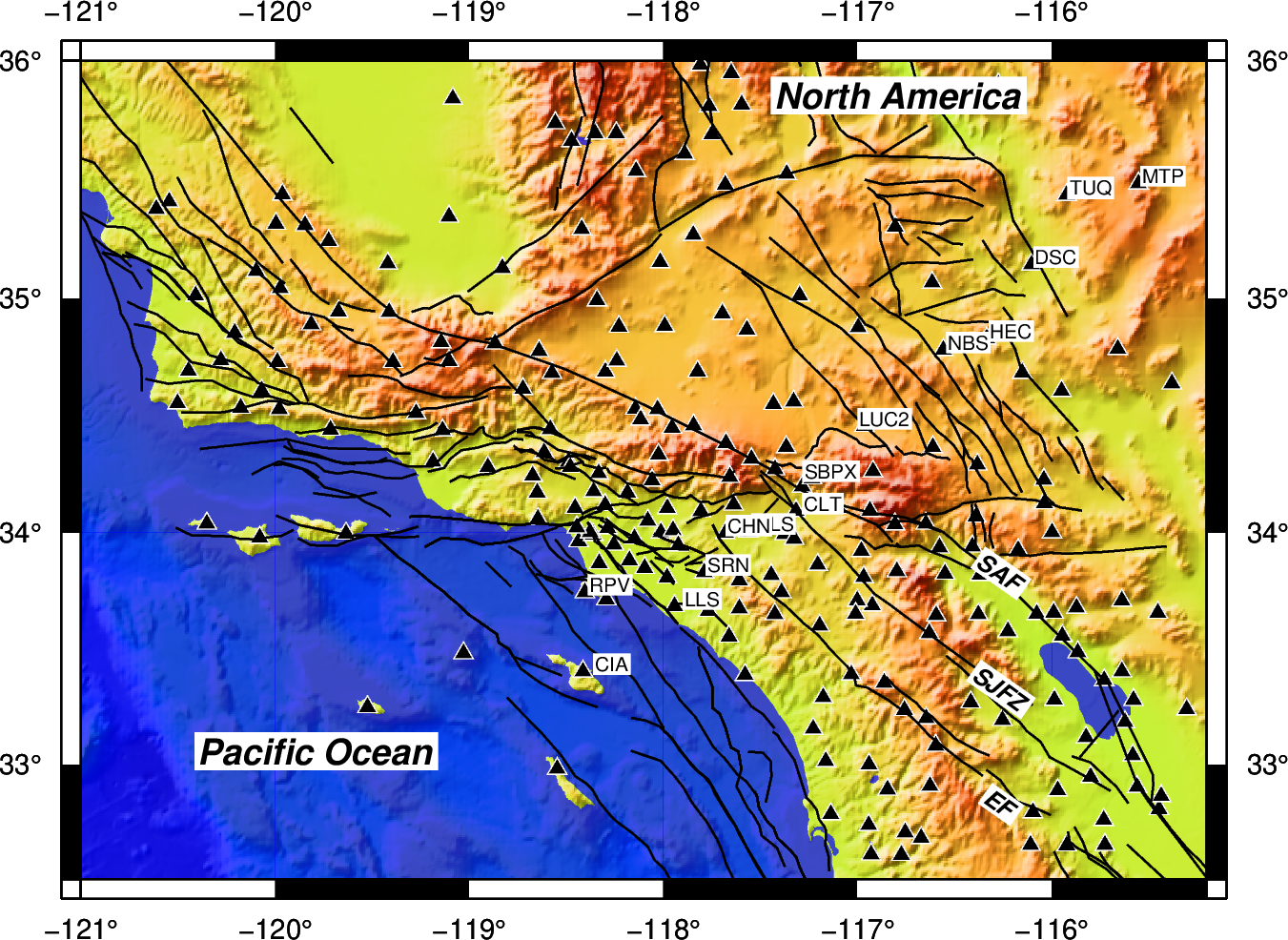}
\caption{The map displays the locations of Southern California stations, denoted by black triangles \citep{Caltech1926}, utilized in this study. Black solid lines represent fault lines in the region. San Andreas Fault (SAF), San Jacinto Fault Zone (SJFZ), and Elsinore Fault (EF) are highlighted. Stations with RFs plotted in Figs. \ref{fig:calirad} and \ref{fig:calitran} are labeled.}
\label{fig:calimap}
\end{figure}

\begin{figure}
\centering
\includegraphics[width=\textwidth]{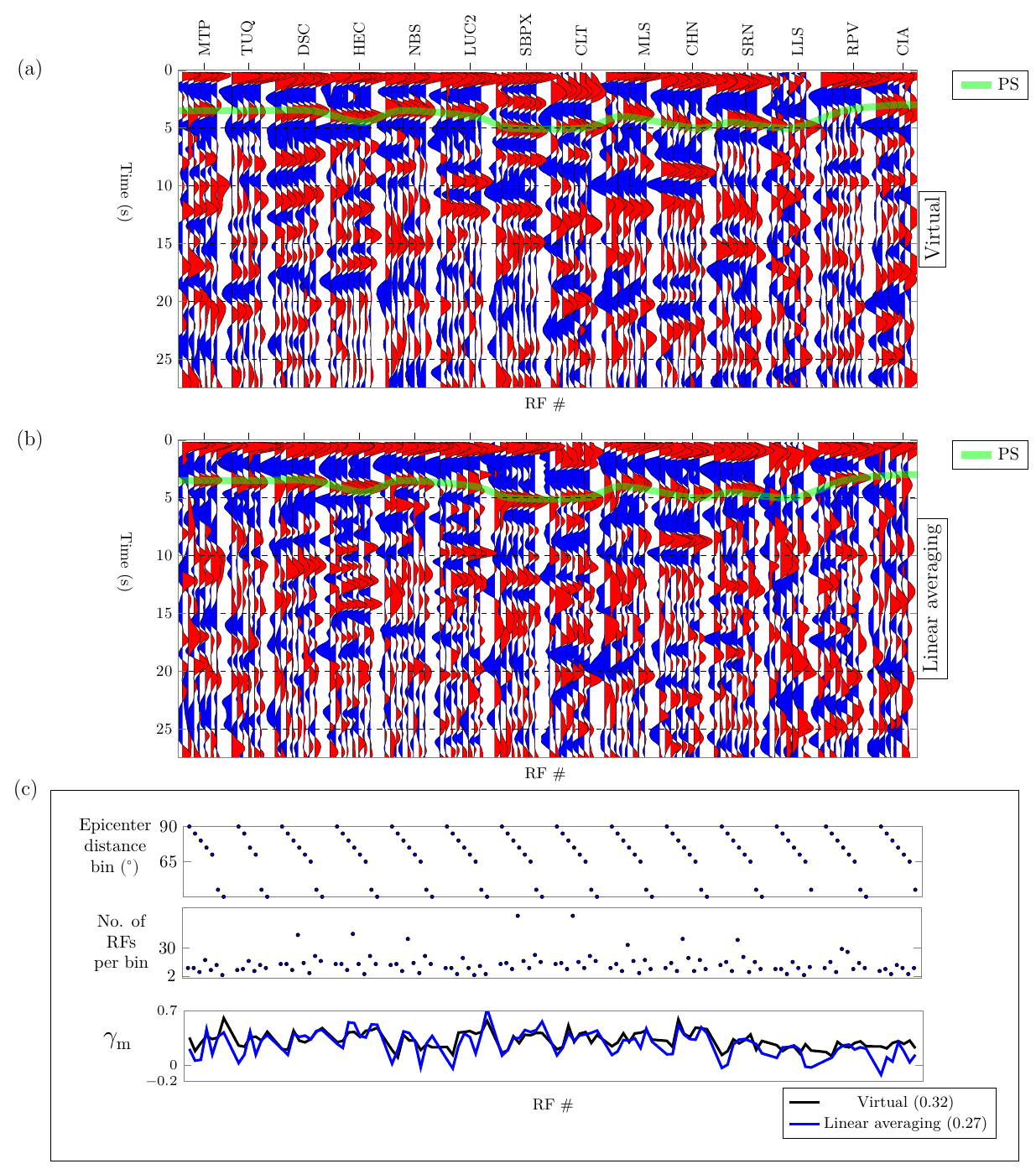}
\caption{(a) Virtual and (b) linearly averaged radial RFs for a station profile transecting the San Andreas Fault (SAF), San Jacinto Fault Zone (SJFZ) and Elsinore Fault (EF) in Southern California (see Fig.~\ref{fig:calimap}). (c) shows epicentral distance bin, number of RFs in the bin and $\ncc_\text{m}$ (mean normalized correlation coefficient) corresponding to RFs in (a) and (b). Converted S-wave (PS), highlighted in green line, shows the variation of Moho depth along the profile. Virtual RFs resemble linearly averaged RFs but demonstrate increased quality across bins and stations, highlighted by higher $\ncc_\text{m}$.}
\label{fig:calirad}
\end{figure}

\begin{figure}
\centering
\includegraphics[width=\textwidth]{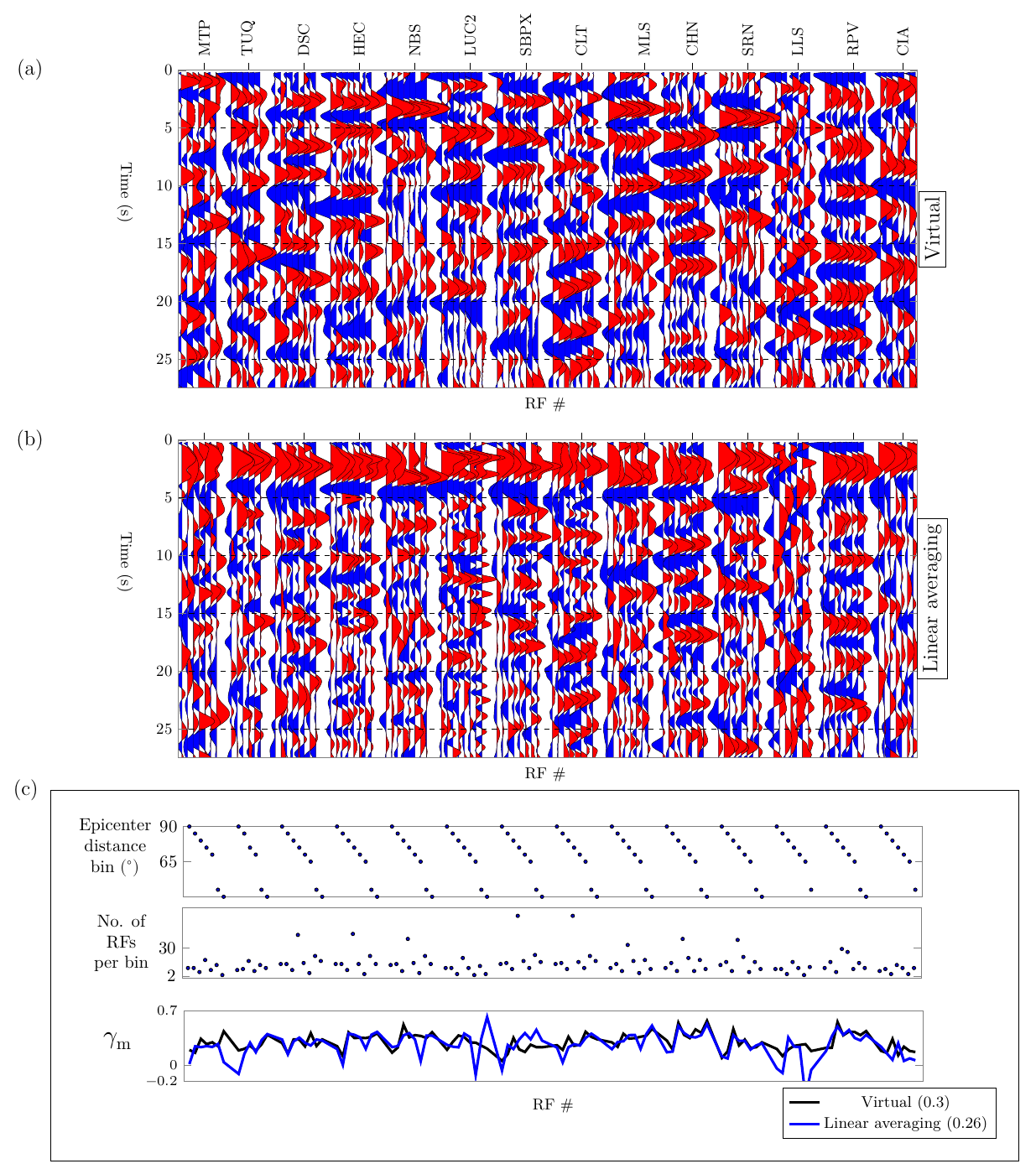}
\caption{Same as in Fig.~\ref{fig:calirad}, but pertaining to tranverse RFs. Note that the high amplitude artifacts at time window 0 to 5s is removed in the virtual RFs.}
\label{fig:calitran}
\end{figure}

Virtual and linear-stacked RFs for the profile that intersects the SAF, SJFZ and EF (refer to Fig.~\ref{fig:calimap}) are presented in Figs.~\ref{fig:calirad} and~\ref{fig:calitran}. In these figures, RFs from the same backazimuth angle bin $304^{\circ}$ to $312^{\circ}$ are plotted to reduce the RF variation due to transverse anisotropy, facilitating the comparison of RF between stations. The radial component of averaged and virtual RFs in Fig.~\ref{fig:calirad} shows the converted S-wave (PS) from the Moho. The PS phase delay time between stations aligns with previous RF studies \citep{Ozacar2009,Ozakin2015}. The PS phase indicates abrupt changes in Moho depth along the profile, most evident at SBPX, possibly due to a vertical offset or sharp dipping Moho \citep{Ozakin2015}. Virtual radial RFs show improved quality compared to averaged RFs, indicated by a higher mean normalized correlation coefficient (MNCC) $\ncc_\text{m}$ between stations. Virtual transverse RFs show significant improvements; a high-amplitude artifact at time window $0$--$5$ s is cleared in virtual RFs. Although interpreting transverse RFs is complex, we emphasize the consistency of RFs between bins and stations, with virtual RFs proving more consistent and of higher quality, as evidenced by the overall higher MNCC.

\begin{figure}
\centering
\includegraphics[width=\textwidth]{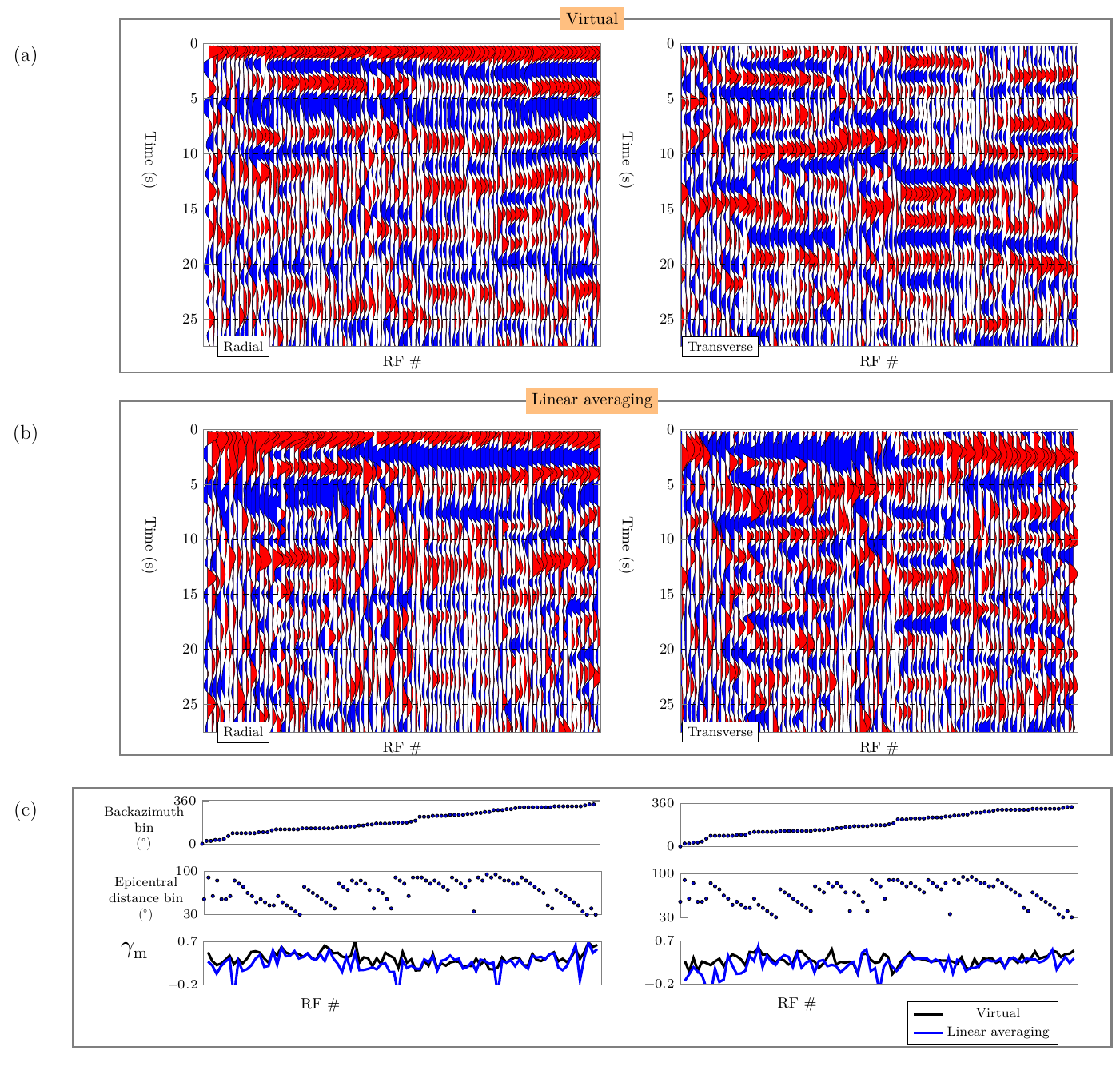}
\caption{Single station (DSC in Fig.~\ref{fig:calimap}) analysis for Southern California case study. (a) virtual and (b) linearly averaged RFs for all RF bins are plotted after ordering based on backazimuth angle. (c) shows the backazimuth angle bin, epicentral distance bin and $\ncc_\text{m}$ (mean normalised correlation coefficient) corresponding to RFs in (a) and (b). The variations in PS arrival times display a single cycle across the entire backazimuth in radial RFs within the $0$--$5$ s time window. Similarly, transverse RFs exhibit one cycle of polarity reversal at the same window, suggesting a dipping Moho.
The enhancement in quality of 
virtual RFs compared to linearly averaged RFs, 
is highlighted by higher $\ncc_\text{m}$ value.
}
\label{fig:dsc}
\end{figure}

\begin{figure}
\centering
\includegraphics[width=\textwidth]{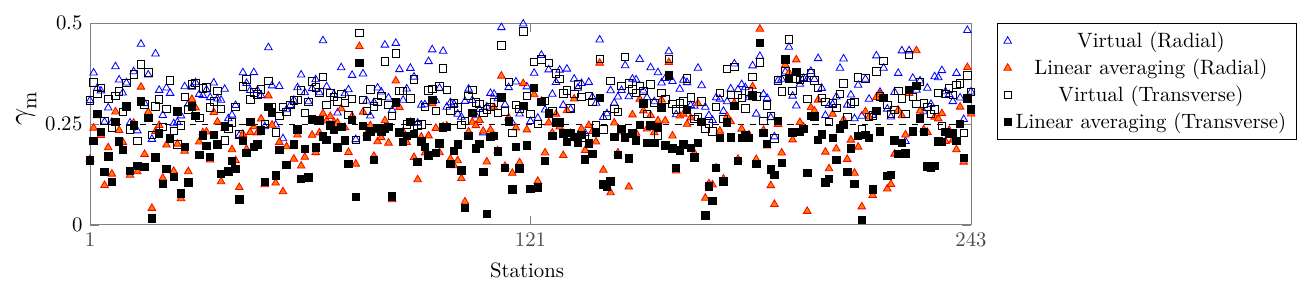}
\caption{Same as Fig.~\ref{fig:cascrr}, but for Southern California stations.
The enhancement in RF quality for both Cascadia and California stations, without altering network hyperparameters, is promising and indicates our network's scalability to various geological regions. Details of the correlation metric ($\ncc_\text{m}$) calculation are in Appendix~\ref{sec:pcc}.}
\label{fig:calicrr}
\end{figure}

To examine the variation of RFs across the entire backazimuth, we plotted RFs from the DSC station in Fig.~\ref{fig:dsc}. SymVAE-generated virtual RFs demonstrate clear and smooth variations along the backazimuth angle, exceeding linear averaged RFs. Similarly to the Cascadia results, seismic signals are more pronounced in virtual RFs than in averaged RFs, evident from the higher MNCC. Virtual radial RFs show a smooth PS delay time variation within a $0$–$5$ s interval, while transverse RFs exhibit polarity reversal in the same interval. Numerical modeling indicates that a dipping Moho, without notable crustal anisotropy, results in a first-order harmonic (single cycle) in the PS arrival time for radial RFs across the entire backazimuth, as well as similar polarity reversal in transverse RFs~\citep{Wang2020}. Visual inspection reveals signs of a dipping Moho in the radial and transverse virtual RFs. However, a detailed RF analysis is beyond the scope of this paper. Lastly, Fig.~\ref{fig:calicrr} shows that the average station-wise MNCC for virtual RFs is generally higher than those for averaged RFs, highlighting the scalability and consistent performance of our method.

\subsection{Sanity checks}
It is important to assess the quality and effectiveness of signals generated by networks for accuracy, reliability, and real-world applicability. We perform the following sanity checks on the virtual RFs to ensure the validity and reliability of our methodology. 

\begin{enumerate}

 \item  SymVAE independently extracts coherent crustal effects from each data point, indicating that there can be substantial variations in crustal effects between any two data points. However, the virtual RFs from nearby stations (PFB and PTRF) in Cascadia exhibit strong similarity. Moreover, the physics governing wave propagation indicates that gradual changes in crustal effects with backazimuth and epicentral distance, as demonstrated in the synthetic RFs in Fig.~\ref{fig:firstplot}, are predictable. Consistent with this, as shown in Fig.~\ref{fig:casrad} and Fig.~\ref{fig:castran}, the virtual RFs in various epicentral bins at all stations exhibit smooth variations. This consistency confirms our method's ability to extract true crustal effects rather than random anomalies.

\item The crustal effects of virtual RFs, generated by various earthquakes that occur at similar locations, are expected to be analogous. To confirm this, we divided the earthquakes in one bin into two distinct groups and train the model separately for each group. After training was performed, the virtual RFs for both groups displayed a remarkable level of similarity, evidenced by a low mean square error.
\end{enumerate}
These checks confirm that the SymVAE results are physically plausible.

\section{Discussion}

Our method has two distinct features: 1) it learns the distribution of nuisance effects without assuming a predefined distribution; 2) by training multiple RFs simultaneously and reducing nuisance effects with a unified network, it does not process each RF independently, compared to most of the existing methods.
Earthquakes usually occur in certain tectonically active areas, causing an uneven spread of teleseismic earthquakes. This uneven distribution complicates the analysis of crustal effects based on backazimuth and epicentral distance, especially for temporary stations that often detect fewer earthquakes, resulting in fewer RFs per bin. Our method allows for the use of all earthquakes, even those with low SNR, potentially enabling the use of several temporary stations.
Refer to the supplementary material for an example illustrating the improvement of backazimuth and epicentral distance coverage.
It is crucial to train temporary stations together with those that record a high number of earthquakes. In addition, incorporating low-magnitude (low SNR) and high-SNR earthquakes is vital, as it enhances the generative capability of the SymVAE network.

Training SymVAE, similar to any deep neural network, necessitates hyperparameter tuning, focusing on coherent and nuisance code lengths. Selecting these optimally aids effectively disentangling nuisance from crustal effects and boosts generative performance. In our case, hyperparameter tuning is carried out through synthetic experiments.

This research assumes that the RFs from nearby earthquakes share coherent crustal effects.
Coherency implies minimal traveltime differences for converted arrivals from such earthquakes. What characterizes minimal and nearby? For a specified maximum frequency of interest, the response depends on the depth of the structure being examined on the receiver's side.
For example, to successfully extract converted seismic waves from the crust or upper mantle, it is possible to group relatively distant earthquakes. 
As travel time differences increase for deeper mantle structures, the earthquakes must be nearer.

\section{Conclusions}
Receiver functions (RFs) are an essential technique for analyzing crust and mantle structures beneath seismic stations; however, they are often contaminated with pseudorandom nuisance effects that reduce the precision and reliability of RF measurements. We introduce an innovative approach to removing these unwanted nuisance effects from RFs through an unsupervised deep learning framework using variational symmetric autoencoders. This method effectively disentangled the nuisance effects from crustal effects in the latent space, allowing the generation of virtual RFs with minimal nuisance effects by optimizing in the latent space. Our approach outperforms linear averaging and phase-weighted averaging in denoising synthetic RFs, which are affected by realistic non-Gaussian nuisance effects, resulting in high-quality and consistent RFs. Unlike averaged RFs, the polarity reversal in transverse RFs along the backazimuth is clearly depicted in virtual RFs. The application of our method in RFs from seismic stations at the Cascadia Subduction Zone's forearc and Southern California has shown strong agreement with previous studies. Our method better delineated the crustal structure which were not as clear in linearly averaged RFs. Virtual RFs demonstrated consistency across backazimuth, epicentral distances, and nearby stations. The method was proved robust by integrating all available RFs, regardless of their signal quality. The automated and human-free process of our method provides both scalability and reproducibility. The compelling results of this study suggest that our approach can greatly improve RF analysis by eliminating nuisances, enabling more precise interpretations of crustal and mantle structures, and revealing new insights previously obscured by noisy RFs, thereby advancing teleseismic imaging and solid-Earth studies.

\textbf{Acknowledgments}
This work is funded by the Science and Engineering Research Board, Department of Science and Technology, India (Grant Number SRG/2021/000205).
This study was enabled by Julia and Python programming languages. Flux package ~\citep{Flux} was used to model and train neural networks. Data were downloaded from IRIS Data Center and Natural Resources Canada Data Center using ObsPyDMT~\citep{Hosseini2017}.

\textbf{Data availability}
Earthquake waveform data used in this study are available in \href{http://service.iris.edu/}{IRIS Data Center}, \href{https://www.earthquakescanada.nrcan.gc.ca}{Natural Resources Canada Data Center} and \href{https://service.scedc.caltech.edu/}{Southern California Data Center}. All codes and packages used in this study are open-access.

\bibliographystyle{unsrtnat}
\bibliography{references_v2}

\appendix

\section{Nuisance Effects In Receiver Functions}
\label{sec:appn1}
\setcounter{equation}{0}
\renewcommand{\theequation}{\thesection.\arabic{equation}}

In this appendix, we show that
the radial receiver function (RF) $r$ from a single teleseismic earthquake can be expressed as follows:
\begin{equation}
    r(t,\p)= \int_{\tau}s_{\text{a}}(t-\tau,\p)\grz(\tau,\p)\text{d}\tau+ \nt(t,\p).
    \label{eqn:A1}
\end{equation}
Here, $s_{\text{a}}$ represents a zero-phase signal, $\grz$ denotes the crustal impulse response, $\p$ is the wavenumber vector, and $\nt$ accounts for noise terms. 
The radial receiver function (RF) is calculated by deconvolving the vertical component seismogram from the radial component seismogram as described in Eq. ~\ref{eqn:one}. 
Fourier transform converts the convolution in the time domain into multiplication in frequency domain. Deconvolution is the process of reversing the effects of convolution and thus can be expressed as a division in the frequency domain.
We express the measured 
seismograms in frequency domain as
\begin{equation}
\begin{split}
  \Dz(\omega,\p) = S(\omega)\Gz(\omega,\p)+\Nz(\omega) \quad \text{and} \\
  \Dr(\omega,\p) = S(\omega)\Gr(\omega,\p)+\Nr(\omega),
  \label{eqn:A2}
  \end{split}
\end{equation}
where $\omega$ is the angular frequency. Then, the radial receiver function is given by
\begin{equation}
    \label{eqn:A3}
    \mathit{R}(\omega,\p) = \frac{\Dr(\omega,\p)\Dz(\omega,\p)^{*}}{|\Dz(\omega,\p)|^2},
 \end{equation}  
where the complex conjugate of the vertical component is $\Dz(\omega,\p)^{*}$.
It is important to recognize that the spectrum of $\Dz$ may contain zeros or very small values, causing instability. To address this issue, various regularization techniques can be employed. We represent the regularization factor by $\lambda$, which varies according to the chosen regularization method, resulting in
\begin{equation}
   \mathit{R}(\omega,\p) = \lambda(\omega,\p)\Dr(\omega,\p)\Dz(\omega,\p)^{*}.
   \label{eqn:A4}
\end{equation}
For example, 
 in the case of water-level regularization ~\citep{clayton1976source},
 \begin{equation}
 \lambda(\omega,\p) = \frac{W(\omega)}{\text{max}\big(|\Dz(\omega,\p)|^{2}, c\;\underset{\forall\,\omega}{\text{max}}\,(|\Dz(\omega,\p)|^{2})\big)},
 \end{equation}
 where $c$ is the water-level parameter and $W(\omega)= \text{exp}\,(-\frac{\omega}{4a^{2}})$ is a Gaussian filter with a width given by the parameter $a$.

By substituting Eq. ~\ref{eqn:A2} into Eq. ~\ref{eqn:A3}, we derive 
 \begin{equation}
 \begin{split}
    R(\omega,\p) = \lambda(\omega,\p)\big(|S(\omega)|^{2}\;\Gr(\omega,\p)\Gz(\omega,\p)^{*}+S(\omega)^{*}\;\Gz(\omega,\p)^{*}\;\Nr(\omega) \\
    + S(\omega)\Gr(\omega,\p)\Nz(\omega)^{*}  +\Nr(\omega)\Nz(\omega)^{*}\big).
    \label{eqn:A5}
    \end{split}
\end{equation}
 In this expression, we have $\Nr(\omega)\Nz(\omega)^{*}=0$, based on the assumption that the seismic noise from the radial and vertical components is uncorrelated. The term $\Gr(\omega,\p)\Gz(\omega,\p)^{*}$ denotes the radial crustal impulse response, whereas $S(\omega)^{*}\,\Gz(\omega,\p)^{*}\,\Nr(\omega)$ and $ S(\omega)\Gr(\omega,\p)\Nz(\omega)^{*}$ account for nuisance from random noise in the seismogram. 
 These nuisance effects are relatively insignificant for high-SNR seismograms but can significantly distort the radial crustal response in low-SNR seismograms. 
 The amplification of seismic noise is influenced by the signature of the source within the regularization factor $\lambda$, which plays a role in the nuisance effects in the RF.
Rewriting Eq. ~\ref{eqn:A5} results in the following equation,
\begin{equation}
    R(\omega,\p)= S_{\text{a}}(\omega,\p)\Grz(\omega,\p)+ \Nt(\omega,\p),
    \label{eqn:A6}
\end{equation}
where
\begin{equation}
\begin{split}
\Grz(\omega,\p)&=\Gr(\omega,\p)\Gz(\omega,\p)^{*},\nonumber \\
S_{\text{a}}(\omega,\p)&=\lambda(\omega,\p)\,|S(\omega)|^{2},\nonumber \\
\Nt(\omega,\p)&=\lambda(\omega)S(\omega)^{*}\;\Gz(\omega,\p)^{*}\;\Nr(\omega)+\lambda(\omega)S(\omega)\Gr(\omega,\p)\Nz(\omega)^{*} \nonumber.
\end{split}
\end{equation}
Importantly, $S_{\text{a}}(\omega)$ represents a zero-phase signal, influenced by both the regularization method and the source signature. In the time domain, Eq. ~\ref{eqn:A6} corresponds to Eq. ~\ref{eqn:A1}.

\section{Coherency in crustal effects}
\label{sec:coherency}
\begin{figure}
\centering
\includegraphics[width=0.9\textwidth]{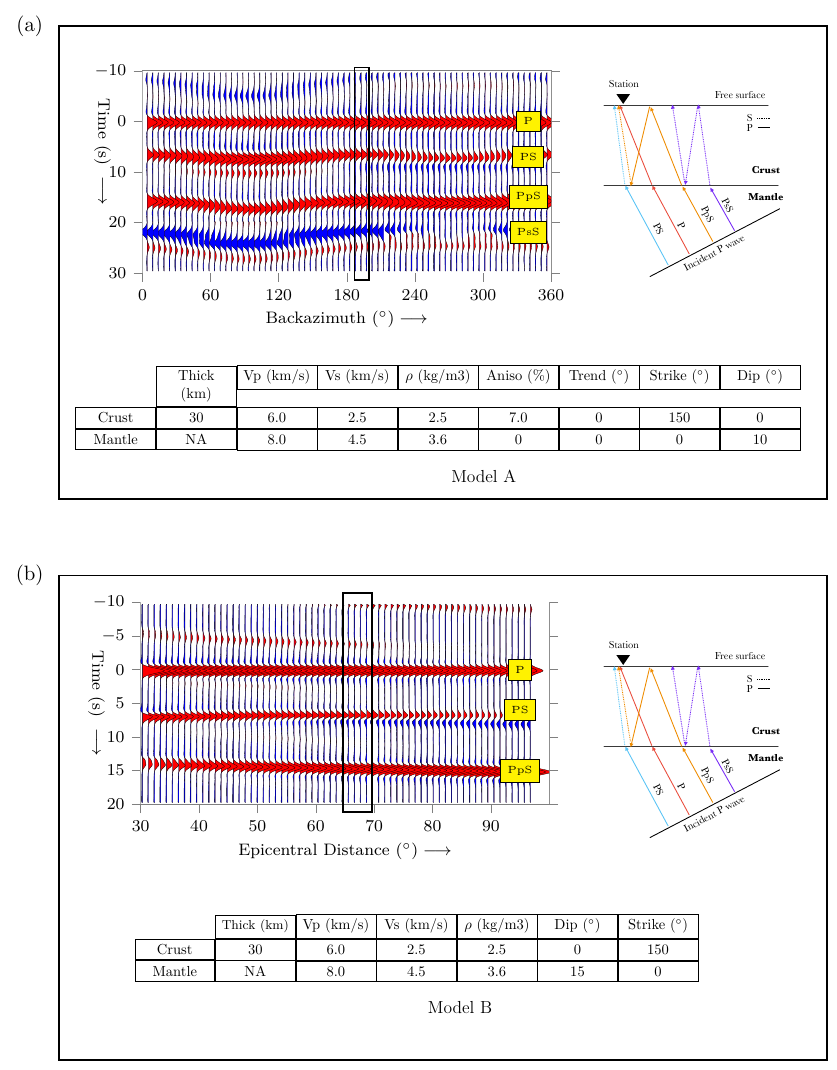}
\caption{ Synthetic radial receiver functions demonstrating the smooth variation with (a) backazimuth and (b) epicentral distance (b).
The two models considered were: (a) one incorporating anisotropy, which allows for variation in receiver functions with backazimuth, and (b) one assuming an isotropic Earth structure.
The seismic phases of interest, generated at the Moho, are marked. This figure illustrates the assumption of coherent crustal effects within bins defined by the backazimuth and epicentral distance. The rectangular boxes indicate the bin widths used in this study. On the left, the raypaths of seismic waves produced at the Moho by an incoming P-wave are shown.}
\label{fig:firstplot}
\end{figure}
A simple numerical validation of coherency in crustal effects across several RFs.

 At crust-mantle boundary, an incoming plane P wave produces S waves through conversion (PS). Moreover, the converted phases could be reflected by the free surface, such as PsS, or when a P-wave is reflected off the free surface, it might transform into S waves, like PpS. Fig.~\ref{fig:firstplot} illustrates some of these crustal phases that compose crustal impulse responses $\gz$ and $\gr$.
To demonstrate the coherence of crustal phases in RFs within a bin, it is necessary to illustrate that both their kinematics and amplitudes vary smoothly with changes in backazimuth and epicentral distance.
Toward this end,
synthetic RFs were produced with the help of the PyRaysum package~\citep{Bloch_Audet_2023}, simulating Earth's crust atop a high-velocity mantle half-space.
To examine how crustal phases change with backazimuth, we first use a crustal model featuring an anisotropic layer with $7\%$ transverse anisotropy and a $10^{\circ}$ dip from the horizontal. The radial RFs for this model are plotted in Fig.~\ref{fig:firstplot}a, which illustrate the changes (due to anisotropy) in the arrival times and amplitudes of crustal phases across different backazimuths.
It is crucial to note that RFs exhibit coherence within a narrow range of backazimuths. Additionally, in the figure, a rectangle highlights the backazimuth bin with a size $10^{\circ}$ used in the synthetic experiment (detailed in Sec. ~\ref{sec:synex}).
Similarly, we now analyze how changes in epicentral distance affect the outcomes. The crustal model shown in Fig.~\ref{fig:firstplot}b features a layer inclined at $15^{\circ}$ from the horizontal and oriented  along N$30^{\circ}$W. This results in changes in the crustal phases PS and PpS as the epicentral distance varies, reflecting the geometric configuration of the crust. Note that when examining a narrow range of epicentral distances (e.g. in an epicentral bin marked using a rectangle), the RFs remain coherent. In conclusion, these simulations illustrate that within a backazimuth epicentral-distance bin, the impact of the crust on the RFs shows minimal variation, confirming our assumption.

\section{Variational Autoencoder: A Deep Generative Model}
\label{sec:vae}

Here, we will review variational autoencoder (VAE) and its training procedure. We will clarify how VAE learns a compressed probabilistic data representation and generates new datapoints from the training data distribution. Following this, we introduce SymVAE, as an extension of standard VAE, designed to improve the interpretation of generated virtual data. Note that, in this paper, we used SymVAE to generate virtual receiver functions with minimal nuisance effects.

VAE is a latent variable model, which means it learns to represent the training data using latent (or) hidden variables.
We represent the $i$th datapoint in the training set $\mathbf{X}$ as $\data_i$, and the vector of latent variables using $\lat$.
Generally, $\lat$ has lower dimensions compared to the datapoint $\data_i$.
Using a geophysical analogy, latent variables can be thought of as the model parameters we estimate while solving inverse problems.
VAE inverts each datapoint, i.e. each element of the training dataset, independently. This means that
each datapoint in the training dataset has a specific latent-variable setting associated with it. 
In traditional geophysical inverse modeling, we use the physics of the problem at hand to derive the posterior distribution associated with the model parameters. In the case of VAEs, we perform variational inference, which means, we train a neural network $\mathbf{h}$ to output the posterior distribution $Q(\lat\mid\data_i,\theta)$ associated with a datapoint. 
This network $\mathbf{h}$ with parameters $\theta$ that outputs the posterior for each datapoint is known as the encoder network.
The posterior is modeled as a multivariate normal distribution.

After performing inverse modeling using the encoder network, we employ another network $\mathbf{f}$ to perform the forward modeling step. Specifically, while the encoder maps each datapoint to its latent variable representation (that is, solving an inverse problem), the decoder $\mathbf{f}$ takes a sample latent code and generates the corresponding datapoint (that is, solving a forward problem). 
The decoder network with parameters $\phi$ outputs the mean of the likelihood distribution $\likl(\data_i\mid\lat,\phi)$ as $\mathbf{f}(\hat{\lat},\phi)$, where $\hat{\lat}$ denotes a sample latent code.
The likelihood is modeled as a multivariate normal distribution with a fixed diagonal covariance matrix.
In this way, the encoder–decoder pair establishes a probabilistic mapping between the data space and the latent space.

We present a concise derivation of the loss function for learning the encoder and decoder parameters in a VAE.
Following \cite{bishop2023deep}, we
assume an independently and identically distributed training dataset $\mathbf{X}=\{\data_1,\data_2,\cdots\}$, and start with an optimization of decoder parameters as
\begin{eqnarray}
    \label{eqn:log-opt}
    \tilde{\phi} &=& \arg \max_\phi \,\log\left(\likl(\mathbf{X}\mid\phi)\right) \\ \nonumber
    &=& \arg \max_\phi \,\sum_i \log\left(\likl(\data_i\mid\phi)\right) \\ \nonumber
    &=& \arg \max_\phi \left[
    \sum_i {\int \post_i(\lat)  \log\left(\frac{\likl(\data_i\mid\lat,\phi)\likl(\lat)}{\post_i(\lat)}\right) \text{d}\lat} + \sum_i D_{\mathrm{KL}}\left({\post_i(\lat)}||{P(\lat\mid\data_i,\phi)}\right)\right],
\end{eqnarray}
where $\post_i(\lat)$ is an arbitrary distribution of the latent variable and
$D_{\mathrm{KL}}(\post_i(\lat)||\likl(\lat\mid\data_i,\phi))$ represents the Kullback-Leibler (KL) divergence between the arbitrary distribution $\post_i(\lat)$ and the posterior $\likl(\lat\mid\data_i,\phi)$. 
As the output of the encoder $\mathbf{h}$ approximates the posterior distribution, we substitute $\post_i(\lat)$ with $\post(\lat\mid\data_i,\theta)$ in the above equation and ignore the calculation of the KL term, which is intractable. In other words, when the encoder output $Q(\lat\mid\data_i,\theta)$ successfully approximates the true posterior $P(\lat\mid\data_i,\phi)$, the KL term approaches zero.
Subsequent to this substitution, we aim to optimize the Evidence Lower Bound (ELBO) to derive the encoder and decoder parameters as
\begin{eqnarray}
    \label{eqn:log-opt}
    \tilde{\phi},~\tilde{\theta} &=& \arg \max_{\phi, \theta}
    \sum_i {\int \post(\lat\mid\data_i,\theta)  \log\left(\frac{\likl(\data_i\mid\lat,\phi)\likl(\lat)}{\post(\lat\mid\data_i,\theta)}\right) \text{d}\lat}, \\ \nonumber
    &=& \arg \max_\phi \left[
    \sum_i \int \post(\lat\mid\data_i,\theta) \log\left(\likl(\data\mid\lat,\phi)\right) \mathrm{d}\lat - \sum_i D_{\mathrm{KL}}\left(\post(\lat\mid\data_i,\theta)||\likl(\lat)\right)\right].
\end{eqnarray}
Here, the prior distribution $P(\lat)$ 
is simply modeled as a multivariate normal distribution with zero mean and identity covariance matrix.

Finally, 
after training VAE, we can select a sample $\hat{\lat}$ from the latent space, referred to in this article as latent code, and generate a new datapoint using the decoder as $\mathbf{f}(\hat{\lat},\phi)$.
In the context of SymVAE, each datapoint $\data_i$ comprises of a group of receiver function (shown in Eq.~\ref{eqn:traindata}).
The latent variable $\lat$ is structured to include $\coh$ and $\nuis^i$ corresponding to the $i$th receiver function in the datapoint.
This structuring of the latent space allows for an interpretable generation of new receiver functions.

\section{Normalized correlation coefficient}
\label{sec:pcc}
\setcounter{equation}{0}
\renewcommand{\theequation}{\thesection.\arabic{equation}}

In this appendix, we demonstrate the computation of the normalized correlation coefficient (NCC), which is used to evaluate the quality of SymVAE-generated and averaged RF.
\subsection{Evaluation of quality in synthetic experiments}
\label{sec:syn-eval}
In the synthetic experiments, the quality of nuisance-minimized RF is determined by comparing it with the true RF. The similarity between the two is quantified by calculating the normalized correlation coefficient (NCC) between the nuisance-minimized RF, $\drec$, and the true RF, $\rrec$, at each data point $\rec_j$ as follows
\begin{equation}
    \ncc(\drec,\rrec)= \frac{{(\drec)}^\mathrm{T} \rrec}{||\drec||_{2}\,\, ||\rrec||_2},
\end{equation}
where $||\drec ||_2$ and $||\rrec ||_2$ denotes the $\ell_2$ norm of  $\drec$ and $\rrec$ respectively.

\subsection{Evaluation of quality in real data application}
\label{sec:real-eval}
In real data applications, since there is no true RF, we calculate the mean normalized correlation coefficient (MNCC) of nuisance-minimized RF with the raw RFs in a datapoint. For instance, the MNCC of the optimized virtual RF $\vir^{\text{opt}}_j$, related to the datapoint $\rec_j$, is given by
\begin{equation}
\label{eqn:ncc}
     \ncc_{\text{m}}(\vir^{\text{opt}}_j,\rec_j)=\frac{1}{n_j}\sum^{n_j}_{i} \ncc\,(\vir^{\text{opt}}_j,\rec^{i}_j),
\end{equation}
where $n_j$ denoted the number of RFs in the datapoint $\rec_j$. The MNCC $\ncc_{\text{m}}$ quantifies how well the optimized virtual RF correlates with the observed RFs in $\rec_j$. A higher $\ncc_{\text{m}}$ indicates more effective information extraction from the observed data. However, calculating the MNCC $\ncc_{\text{m}}$ for the linearly averaged RF using Eq.~\ref{eqn:ncc} can be misleading, as self-correlation of one RF within the datapoint $\rec_j$ can skew the result. To address this, we excluded the identical RF from the averaging process in all NCC calculations within a datapoint.  Thus, the MNCC for the linearly averaged RF, related to datapoint \(\rec_j\), is given by
\begin{equation}
\label{eqn:avg}
     \ncc_{\text{m}}(\rec_j)=\frac{1}{n_j}\sum^{n_j}_{i} \ncc \left(\frac{1}{n_{j}-1}\sum^{n_j}_{k\,,k\neq i}\rec^{k}_j\;,\rec^{i}_j\right).
\end{equation}
Note that this custom averaging of RFs is only employed while calculating the MNCC.
Furthermore, the P-wave at $t=0$ can significantly affect the correlation coefficient in most RFs, reducing the sensitivity to crustal effects. To address this, we calculate the NCC using time samples from $t = 2.5$ s  to the end. 
%

\section{Network Architecture and Training}
\label{sec:training}
In this section, we detail the network architecture used for training on the RFs dataset and provide a short overview of the training process.

In our implementation of the convolutional encoder-decoder architecture, we designed the convolutional layers to extract and reconstruct key features from one-dimensional input RFs. The design of the convolutional layers takes inspiration from AlexNet~\citep{NIPS2012_c399862d}. In both encoders, we employed a sequence of five 1D convolutional layers with the following configuration: Conv(50, 1→5) - LeakyReLU - Conv(20, 5→10) - LeakyReLU - MeanPool(4) - Conv(5, 10→50) - LeakyReLU - Conv(5, 50→40) - LeakyReLU - MeanPool(4). After the last MeanPool, we apply a flattening operation followed by a dense layer that maps the resulting feature vector to the latent space. In decoder, two dense layers are used to transform the concatenated latent code back to the original signal shape. The resulting signal is then processed through five 1D convolutional layers: Conv(50, 1→40) - LeakyReLU - Conv(20, 40→10) - LeakyReLU - Conv(20, 10→10) - LeakyReLU - Conv(5, 10→5) - LeakyReLU - Conv(5, 5→1) - LeakyReLU.

\subsection{Network training}

Datapoint refers to a set of RFs with similar backazimuth angle and epicentral distance. Training dataset consists of all datapoints across stations. For both synthetic and real data, each datapoint was partitioned so that 80\% of the RFs were used for training, while the remaining 20\% were reserved for testing. The testing dataset is not involved in the network training, instead, it is utilized to assess the network's capability to generalize to new, unseen RFs. This evaluation is critical to ensure that the network does not overfit the training dataset. The training dataset for the radial and transverse components was trained on SymVAE network independently. The coherent and nuisance code lengths are crucial for SymVAE to extract significant features from RFs and reconstruct them accurately. The reconstruction loss curves for both training and testing datasets guide the determination of the optimal latent code lengths. For synthetic RFs, the coherent code length ($q$) was set to 50, while the nuisance code length ($p$) was set to 20. For real RF data, these lengths were set to 100 and 20, respectively. The training of the SymVAE network was stopped at 190 epochs, where the reconstruction loss of both the training and testing dataset converged. The learning rate was reduced in three stages: 0.0004, 0.0002, and 0.0001.

The training time of SymVAE for both synthetic and real training datasets is around 15 min. NVIDIA GeForce 3090 (24 GB) GPU card was used in the training. 

%

\end{document}


\maketitle
\begin{center}
\underline{\textbf{\huge Supplementary material}}    
\end{center}

\section{Results from Synthetic Experiment}
\begin{figure}
     \includegraphics[width=\textwidth]{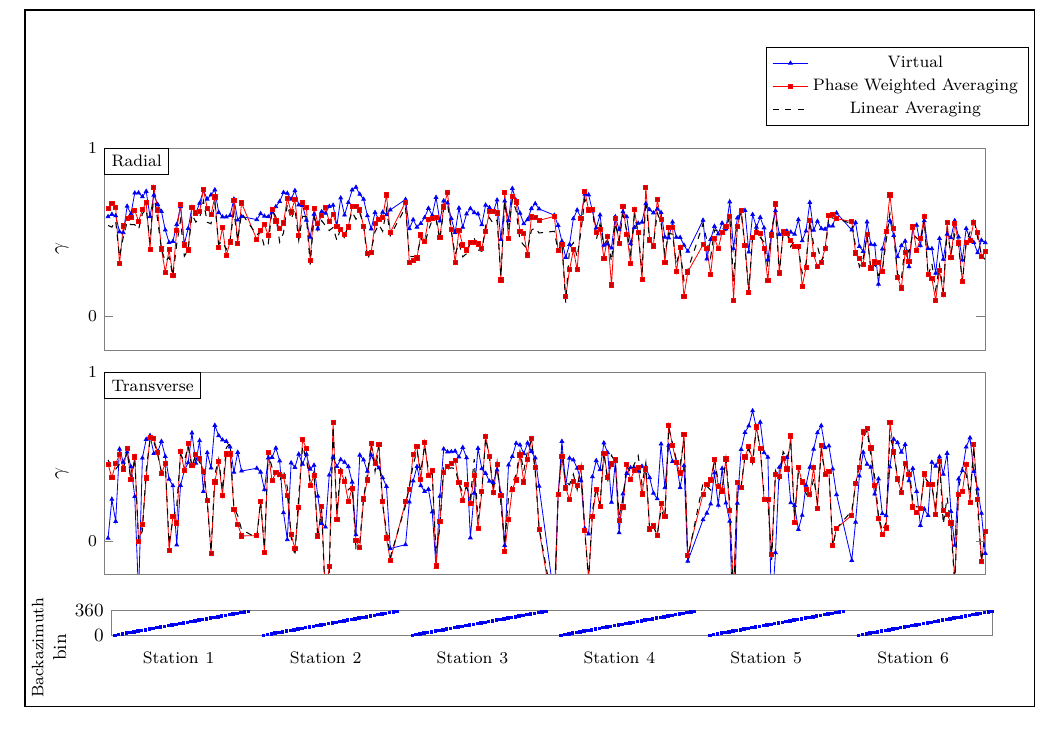}
\caption{Normalized correlation coefficient (NCC) $\ncc$ of virtual, linearly averaged and phase-weighted-averaged RFs w.r.t true RFs for all stations. In radial component, the virtual RFs shows consistent higher correlation compared to averaging techniques.}
\label{fig:AllMSEM1}
\end{figure}

In this section, we show the virtual, linear averaged, phase-weighted averaging, and true RFs for all 6 synthetic stations. Virtual RFs exhibit higher quality compared to averaging techniques, evidenced by higher correlation w.r.t. the true RFs (see Fig.~\ref{fig:AllMSEM1}). We observe a decrease in the correlation as the depth of the initial crustal interface grows; for instance, surface multiples beyond 10 s are not effectively obtained at Stations 4, 5, and 6. This is due to the increase in depth of the crustal interface, which results in greater variability of the converted phases within a bin, thereby causing a loss of coherency.
\vspace{2cm}
\begin{figure}
\centering
\includegraphics[width=\textwidth]{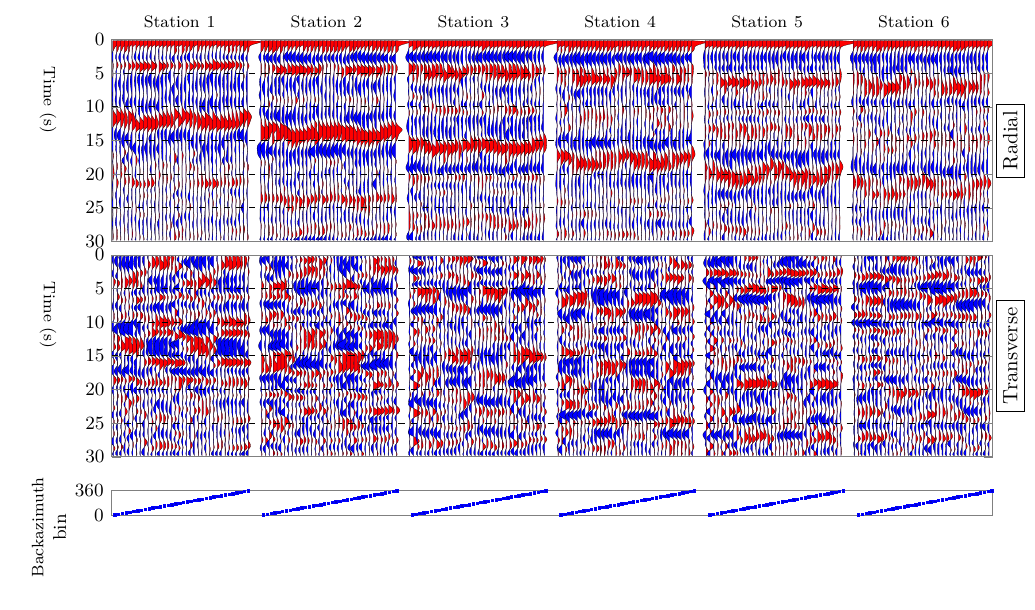}
\caption{ Virtual RFs for all stations (synthetic crustal model). All RFs are from same epicentral bin ($50^{\circ}-60^{\circ}$) across all backazimuth bins.}
\label{fig:AllStaM1}
\end{figure}

\begin{figure}
\centering
\includegraphics[width=\textwidth]{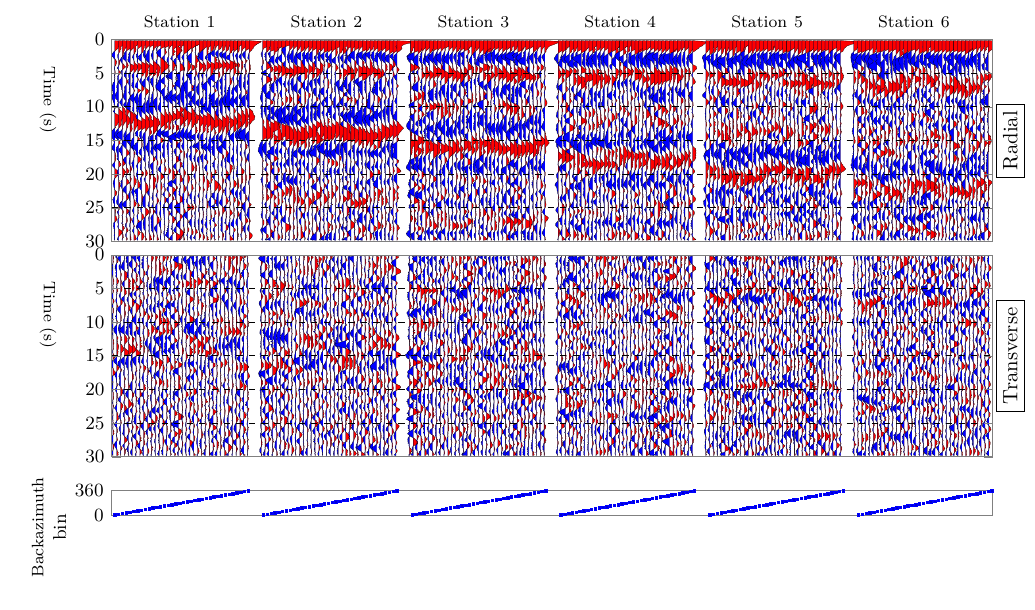}
\caption{ Bin-wise linear averaging of  RFs for all stations (synthetic crustal model). All RFs are from same epicentral bin ($50^{\circ}-60^{\circ}$), across all backazimuth bins.}
\end{figure}

\begin{figure}
\centering
\includegraphics[width=\textwidth]{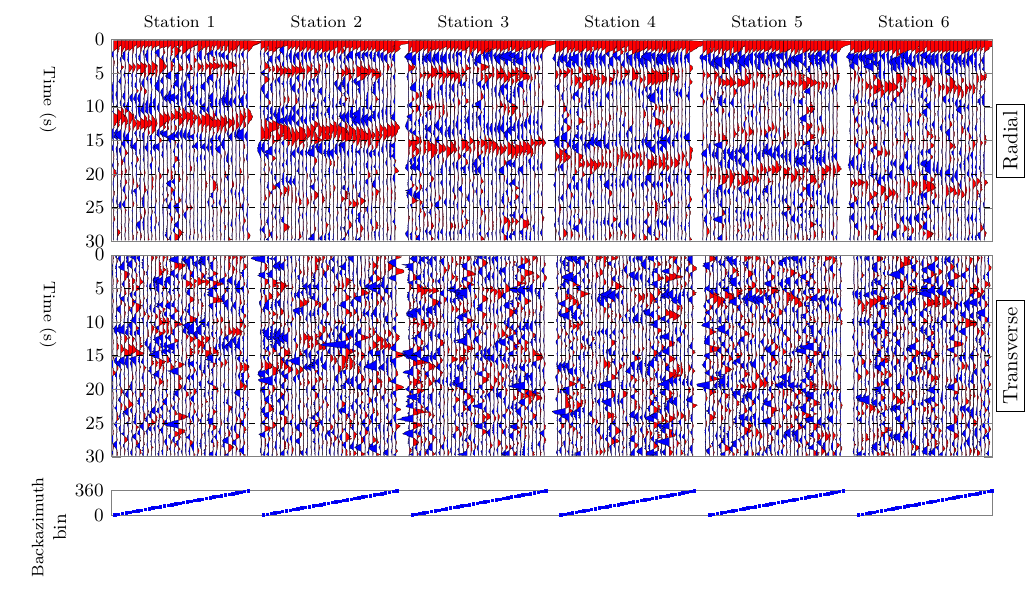}
\caption{Bin-wise phase-weighted-averaged  RFs for all stations (synthetic crustal model). Here, the power of phase averaging $\nu$ is 0.8. All RFs are from same epicentral bin ($50^{\circ}-60^{\circ}$), across all backazimuth bins.}
\end{figure}

\begin{figure}
\centering
\includegraphics[width=\textwidth]{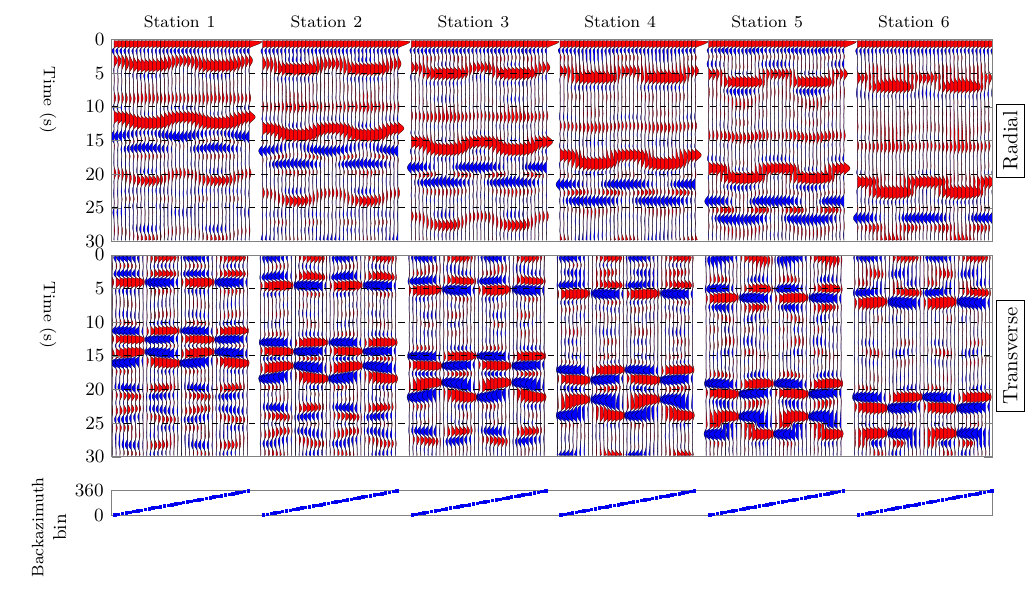}
\caption{True RFs for all stations (synthetic crustal model). All RFs are from same epicentral bin ($50^{\circ}-60^{\circ}$), across all backazimuth bins.}
\end{figure}

\begin{figure}

\includegraphics[trim={0.7cm 0 0.3cm 0},clip,scale=0.7]{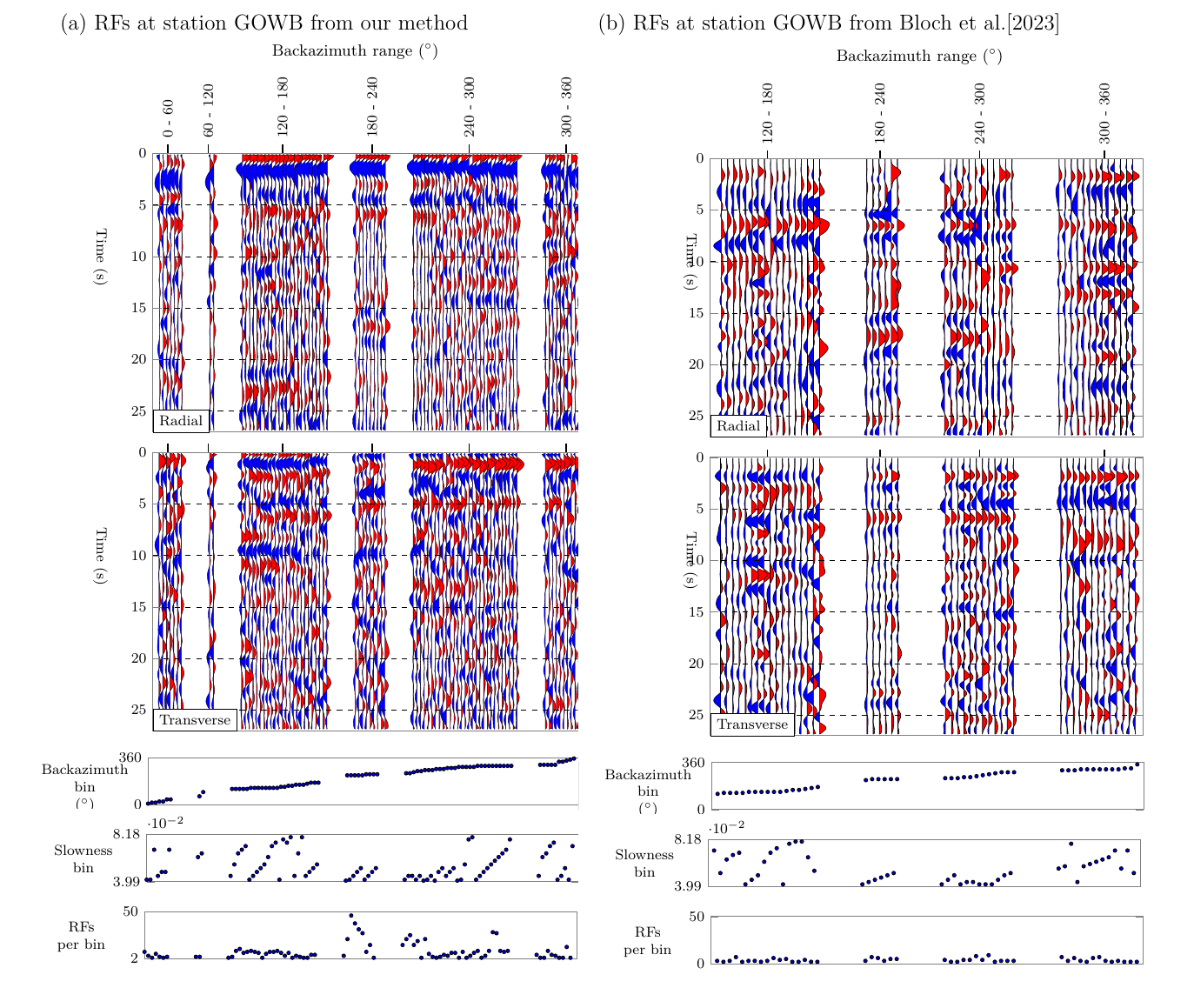}

\caption{Radial and transverse RFs at station GOWB from our method and \textit{Bloch et al. [2023]} demonstrating the increased coverage RFs across backazimuth angle. RFs are sorted by increasing backazimuth and slowness. Groups of RFs within a $60^{\circ}$ backazimuth range are shown together. (a) Our method yields a larger coverage of RFs in terms of backazimuth, relative to (b) \textit{Bloch et al. [2023]}. Notably, the converted S-waves between 0 and 10 s are more distinct in (a) as opposed to (b).}

\end{figure}

\begin{figure}

\includegraphics[trim={0.7cm 0 0.3cm 0},clip,scale=0.7]{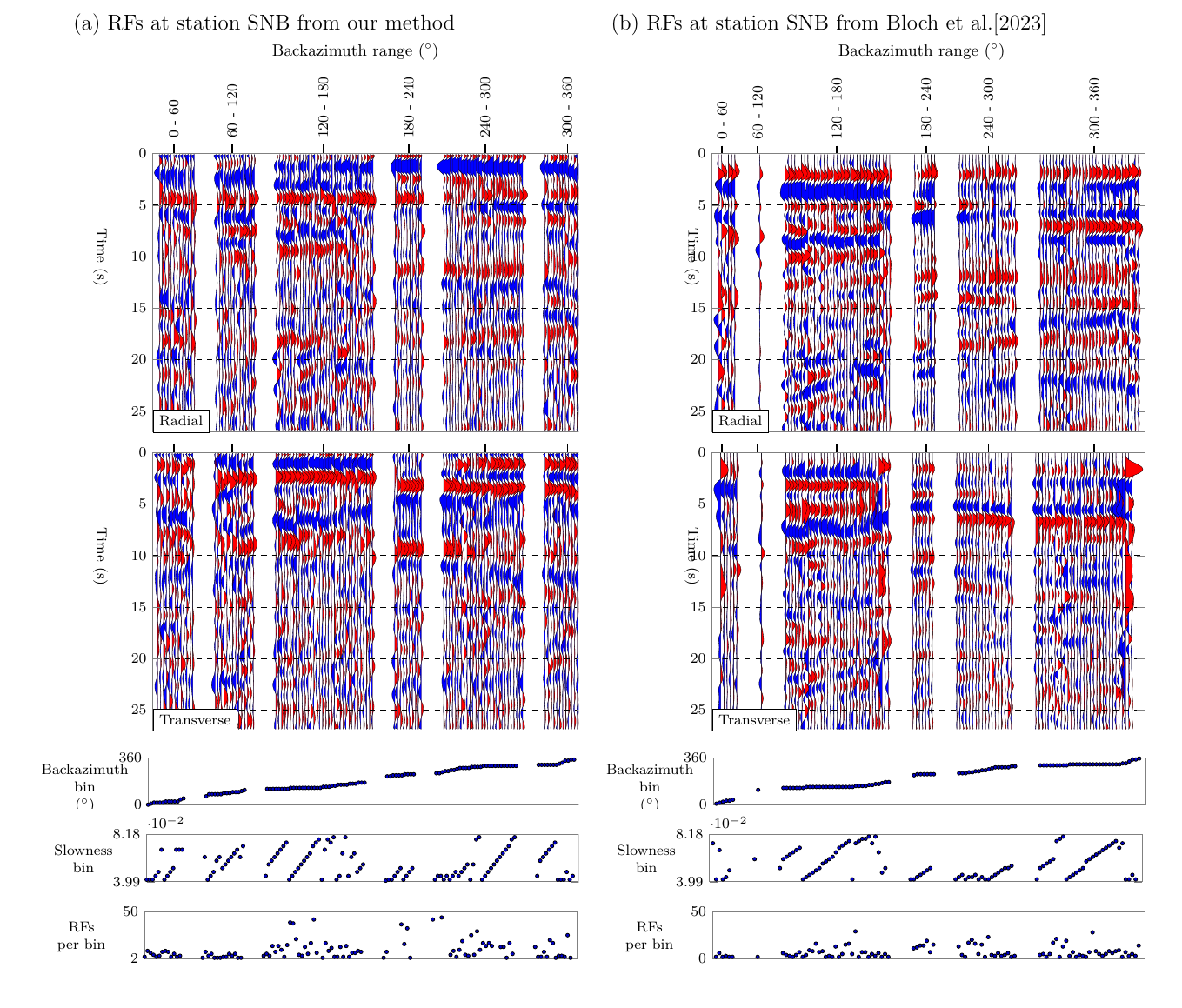}

\caption{Radial and transverse RFs at station SNB from our method and \textit{Bloch et al. [2023]} demonstrating the increased coverage of RFs across backazimuth angle. The RFs are organized by ascending backazimuth and slowness. For ease of comparison, RFs within a $60^{\circ}$ backazimuth range are grouped together. Our method (a) cover RFs for the backazimuth range $60^{\circ}$--$120^{\circ}$, which are missing in (b) \textit{Bloch et al. [2023]}. Converted S-waves between 0 to 10s time window are more prominent in (a) compared to (b). Additionally, notable polarity reversals in transverse RFs are observed between 0 to 5 s, approximately at backazimuth $60^{\circ}$ and $180^{\circ}$, more so in (a).}
\end{figure}